\documentclass[manuscript,natbib=true]{acmart}
\renewcommand\footnotetextcopyrightpermission[1]{}
\usepackage{enumitem}
\usepackage{multirow}
\usepackage{subcaption}
\usepackage{pifont}
\newcommand{\cmark}{\ding{51}}

\AtBeginDocument{%
  \providecommand\BibTeX{{%
    \normalfont B\kern-0.5em{\scshape i\kern-0.25em b}\kern-0.8em\TeX}}}

\setcopyright{acmcopyright}
\copyrightyear{2018}
\acmYear{2018}
\acmDOI{10.1145/1122445.1122456}

\usepackage{lipsum}

\begin{document}

\title{Handling Bias in Toxic Speech Detection: A Survey}\thanks{First two authors contributed equally.}


\author{Tanmay Garg}
\affiliation{%
  \institution{IIIT Delhi}
  \country{India}}
\email{tanmay17061@iiitd.ac.in }
\author{Sarah Masud}
\affiliation{%
  \institution{IIIT Delhi}
  \country{India}}
\email{sarahm@iiitd.ac.in}
\author{Tharun Suresh}
\affiliation{%
  \institution{IIIT Delhi}
  \country{India}}
\email{tharun20119@iiitd.ac.in }
\author{Tanmoy Chakraborty}
\affiliation{%
  \institution{IIT Delhi}
  \country{India}}
\email{tanchak@iitd.ac.in}

\renewcommand{\shortauthors}{Tanmay et al.}

\begin{abstract}
Detecting online toxicity has always been a challenge due to its inherent subjectivity. Factors such as the context, geography, socio-political climate, and background of the producers and consumers of the posts play a crucial role in determining if the content can be flagged as toxic. Adoption of automated toxicity detection models in production can thus lead to a sidelining of the various groups they aim to help in the first place. It has piqued researchers' interest in examining unintended biases and their mitigation. Due to the nascent and multi-faceted nature of the work, complete literature is chaotic in its terminologies, techniques, and findings. In this paper, we put together a systematic study of the limitations and challenges of existing methods for mitigating bias in toxicity detection.

We look closely at proposed methods for evaluating and mitigating bias in toxic speech detection. To examine the limitations of existing methods, we also conduct a case study to introduce the concept of \emph{bias shift} due to knowledge-based bias mitigation. The survey concludes with an overview of the critical challenges, research gaps, and future directions. While reducing toxicity on online platforms continues to be an active area of research, a systematic study of various biases and their mitigation strategies will help the research community produce robust and fair models. \footnote{{\color{red} \noindent\textbf{Disclaimer:} This paper includes examples of toxic speech that contain some profane words. These examples are only included for contextual understanding. We tried our best to censor vulgar, offensive, or hateful words. We assert that we do not support these views in any way.}}
\end{abstract}



\maketitle

\section{Introduction}
\label{sec:intro}
Online social networks (OSNs) have enabled a thriving ecosystem where people from diverse backgrounds can share opinions and ideas. Such online forums are safe spaces for many marginalized groups to support each other and gain a sense of community in the face of discrimination. However, anti-social users often use OSNs to harass and intimidate others. Such behavior manifests in content aimed at harming individuals or groups based on personal attributes such as race, gender, and ethnicity. At its extreme, hate speech (a type of toxic speech) can lead to incidents of offline violence \cite{hatespeechelectoralviolence,hatespeechviolence}. It should be noted that such incidents of violence and hate crime do not occur in isolation. Often negative stereotyping and hostility in the real world get heightened during online interactions. Meanwhile, online hate speech and doxing \cite{doi:10.1080/17440572.2019.1591952} promote offline brutality and even genocides. The online and offline propagation of toxic behavior is a vicious cycle. Such behavior often hampers the ability of marginalized groups to share their opinions freely and further isolates them \cite{saha2019prevalence,doi:10.1089/cyber.2022.0009}. Therefore, the real-world impact of harmful online content and its multifaceted nature have galvanized academic and industrial research toward the early detection and mitigation of such content.

\textbf{Toxic speech.} Throughout this survey, the term toxic speech will be used as an umbrella term to refer to any form of content, including but not limited to hate speech, cyberbully, abusive speech, misogyny, sexism, offense, and obscenity. We follow the definition of toxic speech as given by \citet{paper20} -- ``rude, disrespectful or unreasonable language that is likely to make someone leave a discussion" as an umbrella definition. The literature surveyed here is distributed among various forms of toxic speech, known to adopt different and sometimes misleading terms to refer to equivalent classes \cite{fortunalabelstudy}. The subjectivity in defining toxicity can also be attributed to the lack of an in-depth understanding of toxicity through the lens of social and psychological science. Machine learning researchers and practitioners often overlook the root cause of toxic behavior on the Internet \cite{Suler}. Such root cause analysis can help understand the power dynamics of hate \cite{Navarro}, the prevalent discrimination, and its evolution on the Internet \cite{agneta,10.1145/3419249.3420142}. Recent studies in implicit hate speech \cite{toxigen,latent_hate} attempt to bridge this gap; however, such practice in computational methods for toxicity detection is not yet well adopted.

\textbf{Vulnerable groups.} In relation to toxic speech, there is always a group (or an individual from the group) at the receiving end of all the negativity and hate. Such groups can be termed as vulnerable/target/protected groups. For bias evaluation, they may be even referred to as subgroups or identity groups. Each of these groups can be identified with negative phrases that are used to refer to the individuals of the particular group. Such phrases are called ``identity terms''. For example, when analysing hate speech against Jewish people, "k*k*" is a very commonly used slur. 

\textbf{Unique characteristics of online toxic speech.} Toxicity detection is a highly subjective task, and context plays a crucial role in determining whether the content can be flagged as toxic. Different online platforms and legal agencies use varying definitions of toxicity. Meanwhile, existing datasets display a variety of biases via their collection pipeline \cite{paper20,paper19} and are subject to unreliable annotations \cite{paper18}. This paper explores why these characteristics render toxic speech vulnerable to various unintended biases and the methods proposed to handle them.

\begin{figure}[!t]
    \centering
    \includegraphics[width=15cm]{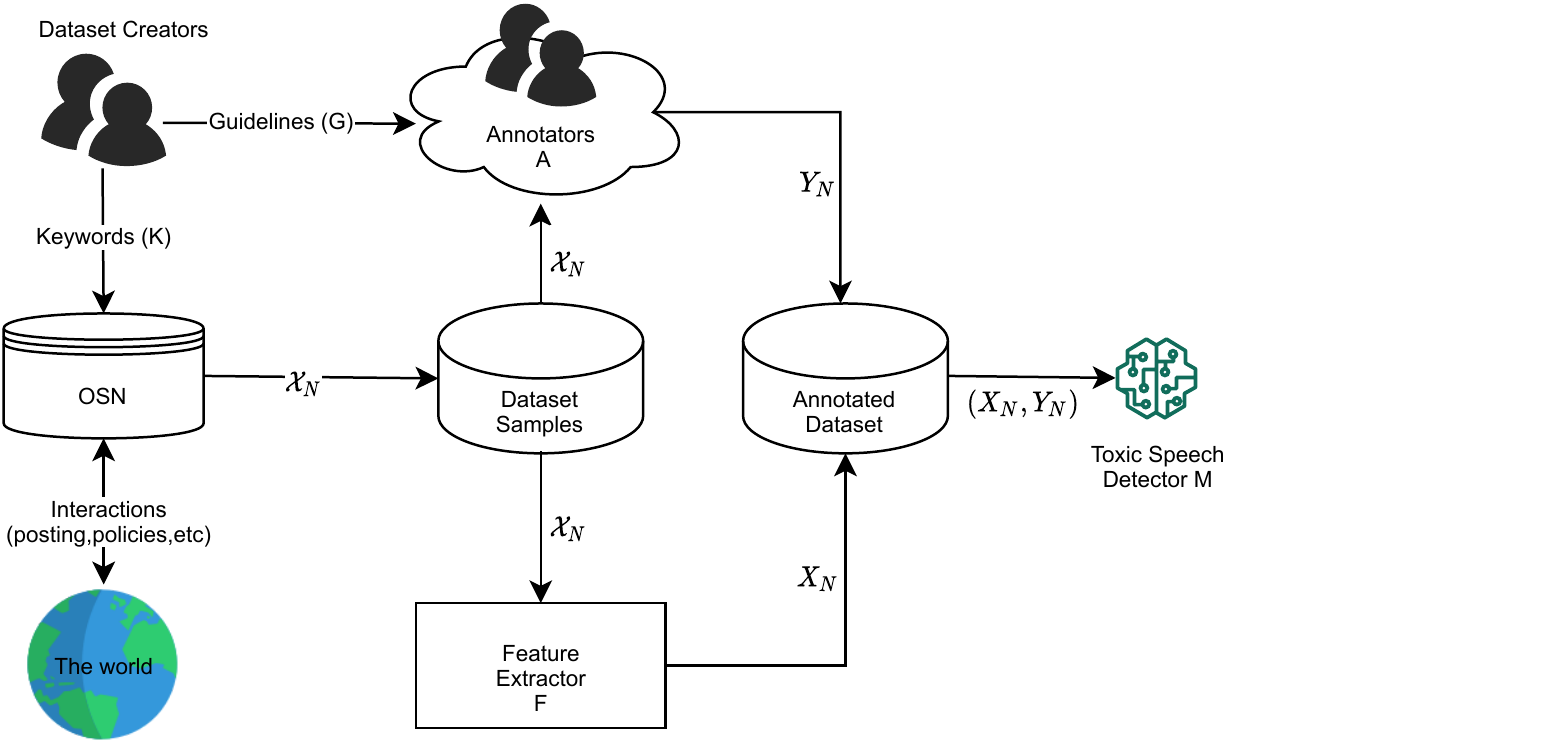}
    \caption{The pipeline of a toxic speech detection model can be visualised as a sequence of data transformations: (i) the OSN sampler API ($S:K\rightarrow \mathcal{X}_N$) takes as input a set of keywords $K$ (possibly empty) and returns a set of samples $\mathcal{X}_N$; (ii) the annotation function ($A:\mathcal{X}_N,G\rightarrow Y_N$) converts $\mathcal{X}_N$ and the annotation guidelines G to a sequence of annotations $Y_N$; (iii) the feature extraction function ($F:\mathcal{X}_N\rightarrow X_N$) converts $\mathcal{X}_N$ to a sequence of feature vectors $X_N$. (iv) Finally, the function ($M:\mathcal{X}_N,Y_N\rightarrow Y'_N$) aims to predict the toxicity label. 
    Biases based on the source of harm are associated with these transformations as (i) sampling bias with $S$, (ii) lexical bias with $(F, A)$, and (iii) annotation bias with $A$. Meanwhile, the biases based on the target of harm manifest when the prediction $Y'_N$ of the model $M$ are analysed.}
    \label{fig:datatransformations}
    \vspace{-4mm}
\end{figure}

\textbf{Scope of the survey.} Bias mitigation methods in natural language processing (NLP) have been extensively studied \cite{GarridoMuoz2021, blodgett-bias-in-nlp-survey, weidinger2021ethical}. 
Consequently, the mitigation techniques in NLP have been helpful in tasks such as textual entailment \cite{textualentailmentbias}, or reading comprehension \cite{readingcomprehensionbias}. It also motivated researchers to apply bias mitigation for toxic speech detection. While the respective methods have progressed, the results are not as effective as expected. The argument goes back to the subjective /objective nature of tasks like toxicity detection vs. textual entailment. It inspired us to conduct a survey that can help better understand the current state-of-the-art bias mitigation in toxicity detection. The scope of our study is not to discuss the comprehensive research of biases in NLP; instead, we present a thorough analysis
of the methods that study \emph{bias as applied to the case of automatic toxicity detection}. Additionally, several surveys have already examined the existing literature in modeling the toxicity detection task \cite{toxicspeechsurvey1,toxicspeechsurvey2}. We do not survey all the existing toxicity detection methods; instead, we focus on a subset of them, exploring and mitigating bias in toxic speech detection. Meanwhile, \citet{generalisable-hs-survey} surveyed the literature addressing the robustness of hate speech detection methods and addressed the subject of bias in hate speech detection. While their discussion remained general commentary, we aim to develop an extensive understanding of these methods. 

To begin with, we develop a taxonomy of bias based on the sources and targets of harm. Each bias mitigation method can be applied to one or more data transformation stages of a machine learning (ML) pipeline \cite{frameworksuresh}. Subsequently, Table \ref{tab:datasets} provides a summary of popular hate speech datasets and their usage for the study of evaluating and mitigating bias in toxic speech detection. As a part of our literature survey, we also review the reproducibility of existing debiasing methods employed for toxicity detection. The details of the experiments are provided in Appendix \ref{sec:rep_research}. Meanwhile, we also discover a compelling phenomenon of bias shift in knowledge generalization-based methods and provide a short description of the same in Appendix \ref{sec:casestudybiasshift}. 

\textbf{Survey methodology.} Following \citet{generalisable-hs-survey}, we considered Google Scholar as the primary search engine to curate relevant papers. We focused on toxicity debiasing studies done within the last five years and mainly looked at research published in $2016$ onwards. We started with relevant keywords such as ``bias'', ``toxic speech'', ``abusive speech'', and ``hate speech'', shortlisting a seed set of papers through their abstracts. Papers were also collected from recent proceedings of relevant data mining, NLP and web-science conferences (ACL, EMNLP, NAACL, AAAI, WebSci, ICWSM, etc.), journals (TACL, TKDD, PLOS, etc.) and workshops (WOAH etc.). We also visited the citing and cited papers of the seed papers to locate relevant papers further. This shortlisting process was done between September-November $2021$.

\begin{table}[!t]
\caption{List of popular toxic speech datasets in the study of bias as discussed in this paper. `E' (Evaluates) indicates when a dataset was used to establish the presence of bias. `M' (Mitigates and Evaluates) indicates if debiasing techniques were performed on the dataset. Here sampling, lexicon, and annotation are sources of bias, whereas race and gender are targets of bias. Note that this table is a representative sample of the literature surveyed here and is not a comprehensive set.}
\label{tab:datasets}
  \begin{tabular}{ | p{2.1cm} |p{1.4cm} | p{0.7cm} | p{0.9cm} | p{1.4cm} | p{1.8cm} | p{1.7cm} | p{0.45cm} | p{1.2cm} | } 
    \hline
    Dataset&Sampling& \multicolumn{2}{c|}{Lexical}& Annotation& \multicolumn{2}{c|}{Racial}& \multicolumn{2}{c|}{Gender}\\
    \cline{2-9}
    &E&E&M&E&E&M&E&M  \\
    \hline
    Founta \cite{founta}&\cite{paper19,paper8}&\cite{paper19}&\cite{paper10}& &\cite{paper5,paper3,paper2}&\cite{paper11,paper10,paper1}&\cite{paper3}&\cite{paper9}  \\
    \hline
    W\&H \cite{waseemandhovy}& \cite{paper19,paper16,paper8} & \cite{paper19} & & \cite{paper17,paper6} & \cite{paper6,paper5} & & & \cite{paper19,paper9}  \\
    \hline
    Wulczyn \cite{wulczyndataset}& \cite{paper19,paper13} & \cite{paper19} & \cite{paper21,paper20} & \cite{paper22,paper23} & & & &  \\
    \hline
    Davidson \cite{davidsondataset}& & &\cite{paper21} &\cite{paper6} &\cite{paper6,paper5,paper1,paper3} &\cite{paper11} &\cite{paper3} &  \\
    \hline
    Waseem \cite{paper17}& & & &\cite{paper17,paper6} &\cite{paper6,paper5} & & &\cite{paper9}  \\
    \hline
    GHC \cite{ghcdataset} & & &\cite{paper7} & & & & &  \\
    \hline
    Stormfront \cite{whitesupremacystormfrontdataset}& & &\cite{paper7} & & & & &  \\
    \hline
    Evalita2018 \cite{evalitadataset}& & & & & & & &\cite{paper12}  \\
    \hline
\end{tabular}
\vspace{-4mm}
\end{table}

\section{Categories of bias}
\label{sec:cat_bias}
All machine learning models, including toxicity detection, assume some bias in the data to perform predictions \cite{paper20}. However, we do not intend the toxicity prediction to vary based on the speaker's racial background, for example. If a model exhibits such bias, we call it {\em unintended bias}. In the rest of the paper, we use `bias' to refer to unintended bias in toxic speech detection.

\textbf{Background on NLP Bias.}
In cognitive science, bias is assumed to be a shortcut (valid/invalid) —- our brain resorts to when informing our actions and interactions with others. Biases can develop due to a limited worldview and repeated exposure to similar behavior in the surrounding \cite{Haselton2015}. Owing to which filter bubbles \cite{10.1145/3383313.3418435} and echo chambers get formed, further amplifying the stereotypes. When such biased interactions are employed to train natural language models, they learn stereotypical statistical associations prevalent on the Internet \cite{caliskan}. The seminal work by \citet{bolukbasi} led the initiative of establishing gender (male vs. female) bias in non-contextual embeddings trained on large-scale web corpora. This work prompted similar studies \cite{caliskan, blodgett-bias-in-nlp-survey} of bias in NLP towards other social/personal attributes.

\citet{blodgett-bias-in-nlp-survey} broadly classified the NLP biases as either allocational or representational harm. \citet{sun-etal-2019-mitigating} further granulated representation harm into denigration, stereotyping, recognition, and under-representation when evaluating various NLP tasks and their associated language models for mitigating gender bias. In a tangential approach, \citet{shah-etal-2020-predictive} and \citet{b-etal-2021-overview} looked at biases in NLP through the lens of data and modeling pipelines.  \citet{shah-etal-2020-predictive} proposed a pipeline to highlight the origin of various biases and their eventual outcome disparity. On the other hand, \citet{b-etal-2021-overview} largely classified the biases as related to either data source, label/annotation, and embedding representation. 

\begin{figure}[!h]
    \centering
    \includegraphics[width=8cm]{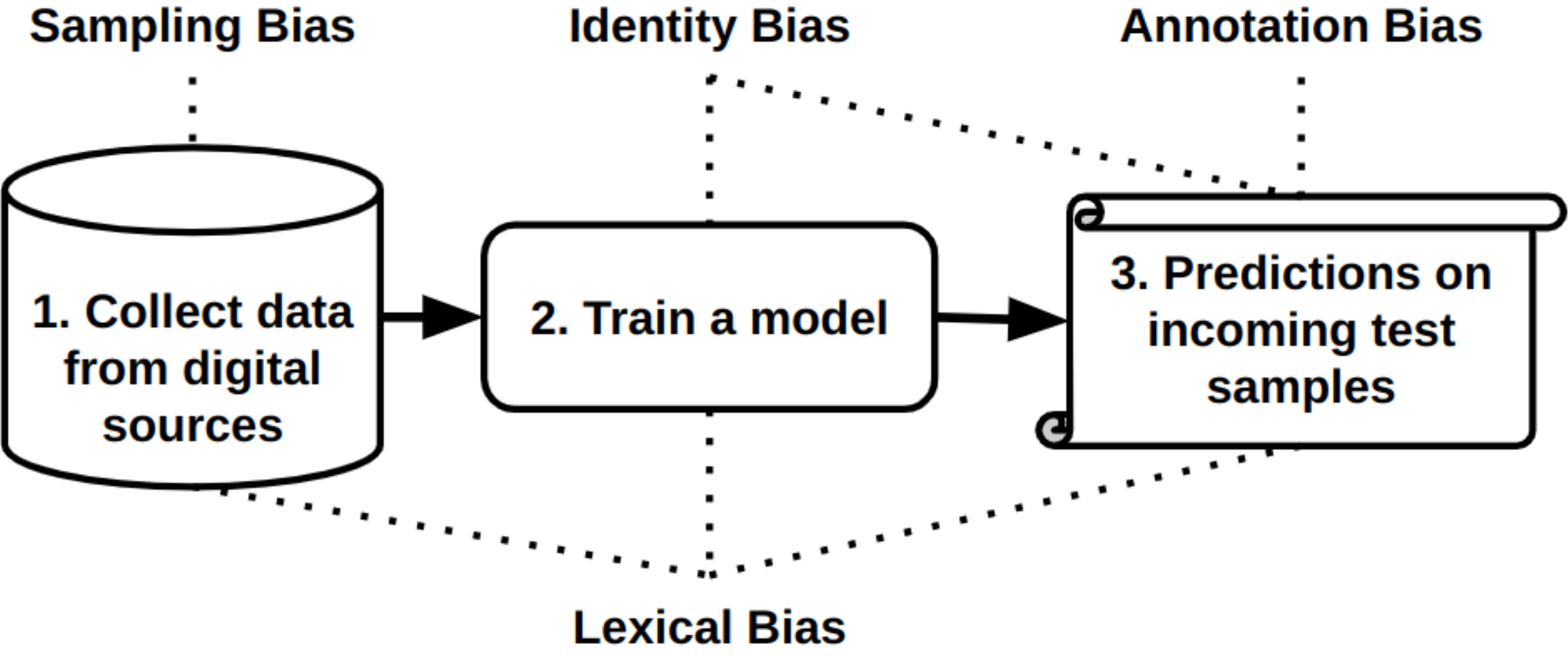}
    \caption{The pipeline of training and evaluation of NLP models can be visualized as a sequence of (i) data collection, (ii) training of models, and (iii) employing the models on incoming test samples. Here collecting data from digital sources can introduce sampling and lexical biases (if specific web pages/digital footprints are ignored). Model training can also introduce linguistic bias when frequently picking co-occurring phrases. Training processes and skewed datasets can introduce identity-based prejudice (race, gender, age, political affiliation, etc.). If predictions on downstream NLP tasks happen on human-labeled gold datasets, it can be a source of annotation and sampling biases (not depicted here). Both linguistic and identity-based biases manifest during the model evaluation and in-the-wild testing.}
    \label{fig:llm_pipeline_bias}
    \vspace{-3mm}
\end{figure}

\textbf{Relation to NLP Bias.} A simple bias mapping pipeline for training a generic NLP model is depicted in Figure \ref{fig:llm_pipeline_bias}. Comparing the training pipelines in Figures \ref{fig:datatransformations} and \ref{fig:llm_pipeline_bias}, we can observe how the first step (data collection and sample) in both workflows is primarily the source of sampling bias. In the case of toxicity detection, biased sampling is performed to gain a higher percentage of toxic samples. Consequently, in broader NLP modeling, it occurs due to a skewed ratio in terms of quality and quantity of digital footprints of specific topics (such as content in support of LGTBQ vs. against them). Sometimes, the training of NLP models can be unsupervised, in such cases, the annotation bias in the pipeline shifts to the last stage when it is employed on downstream tasks that use gold-label datasets to establish performance. All the shortcomings of human annotation apply to various NLP tasks. Some are more static and objective, leading to less bias and disagreements. Meanwhile, others can be highly subjective and ephemeral. As a result of various sampling and annotation biases, we observe the prevalence of identity-based harm introduced due to spurious lexical correlations. We term the bias/prejudice against a person's demographic or psychographic characteristic or any identifiable physical or mental characteristic as {\em identity bias}. As such, prejudice against specific identity groups can be analyzed separately based on the target of harm. Inspired by various approaches, we develop the following taxonomy based on the -- (i) sources and (ii) targets of harm. Biases studied in this survey are not unique to the task of toxicity detection; instead, they are an extension of the biases in NLP applications.

\textbf{Based on {\em sources} of harm}:
We take inspiration from \citet{frameworksuresh} to group the surveyed methods into categories based on the source of downstream harms during the data collection process. The authors defined the process, consisting of -- selecting a \textit{population}, selecting and measuring \textit{features}, and \textit{labels} to use. We study categories of bias according to the transformations related to these steps (described in Figure \ref{fig:datatransformations}) as -- sampling , lexical, and annotation bias.

\begin{figure}[!ht]
    \centering
    \includegraphics[width=12cm]{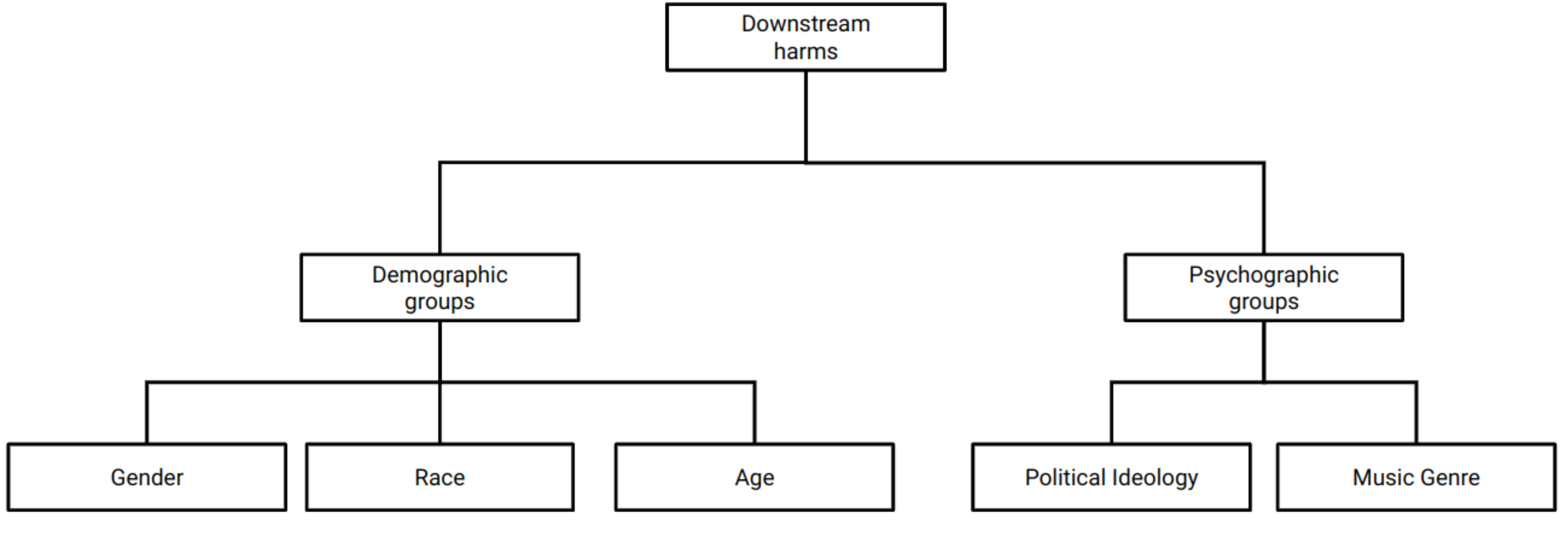}
    \caption{A taxonomy of bias based on the downstream harm. The harms can be inflicted through the real-world deployment of a toxic speech detection system.}
    \label{fig:downstream-harms}
    \vspace{-4mm}
\end{figure}

\textbf{Based on {\em targets} of harm}: The next three categories of bias in toxic speech are each dedicated to a target group of downstream harm (Figure \ref{fig:downstream-harms} ) -- (i) racial bias, (ii) gender bias, and (iii) psychographic bias like political affiliations \cite{Huszr2021}. However, the study of biases based on psychographic attributes (grouping individuals w.r.t their beliefs and interests) is yet to gain popularity. Through this survey, we hope to encourage future exploration of bias categories tied with psychographic attributes.

\textbf{Other categorizations of bias.}
Some of the papers surveyed here present implicit vs. explicit bias categorization. We refer the readers to \cite{mitigating-implicit-bias-2021} for an understanding of this bias stratification. \citet{blodgett-bias-in-nlp-survey} use a taxonomy that categorises between \emph{allocational} and \emph{representational} harms. While this taxonomy is useful to segment existing papers based on their motivation of handling bias, the taxonomy we developed lets us approach the proposed methodologies from an application point of view.

\section{Biases based on source of harm}
\textbf{Overview.} This section will cover  three main categories of biases that stem from the data processing step. We begin our discussion with sampling and annotation bias and then cover lexical bias as the consequence of sampling and annotation. Sampling bias occurs at the first step of data collection. The workaround employed to obtain a more significant percentage of toxic comments against vulnerable communities eventually manifests as bias against the very community it intends to protect. 

Such a dataset may require fine-grained labels obtained at large-scale by human-aided annotations. However, performing annotations for subjective tasks such as labeling toxicity is tricky because there is no standard definition of what can be considered toxic. Here the cognizance of the annotators about the existing discrimination and current socio-political status quo is essential in obtaining valuable annotations. Especially when looking into toxicity against a particularly vulnerable group such as sexual orientation, it is necessary to consider the annotators' political, religious, and social views before annotations. It should be noted that such social beliefs need not be independently formed. Religion, for example, is not a personal belief system but can be culturally reinforced, sometimes acting as a proxy for race and ethnicity.  

Building further on the context of sexual orientation, if the annotators are provided with meta-data and background of the speakers participating in the comments, it can be a double-edged sword. On the one hand, it can provide the context to correctly label such comments as non-toxic despite containing seemingly offensive terms. On the other hand, it can trigger latent biases in the annotators. They can willfully provide erroneous labels and label the interaction as hateful and harmful, leading to suppression of the voices of the minority. 

Once the data is collected and annotated, it is ready for pre-processing and modeling. Once modeled, we can retrospectively analyze the models for downstream harms and biases it incorporates against different identity groups like race, gender, or both. One such bias is lexical. Lexical bias can be evaluated at both dataset (source of harm) and modeling level (downstream harm). For this survey, we include it with the source of harm as lexical biases may or may not be identity specific and can be generally spurred from the presence/absence of words in both toxic and non-toxic labels. It should be noted that all biases we discuss in the coming sections, i.e., sampling, annotation, and lexical biases, directly/indirectly reflect the varying linguistic and social attributes and subjective nature of toxic content. An abbreviation of various terms used in  our discussion of biases is enlisted in Table \ref{tab:termsused}. A complete description of the various bias metrics and their usage is provided in Appendix \ref{sec:evaluationmetrics}.

\begin{table}[!h]
\caption{Abbreviations and expansions used throughout the paper.}
\label{tab:termsused}
\scalebox{0.8}{
  \begin{tabular}{ | l | l |}
    \hline
    Abbreviation& Expansion\\
    \hline
    SBET&Synthetic Bias Evaluation set\\
    BSW&Bias Sensitive Words\\
    SOC&Sampling and Occlusion\\
    $L_{CE}$&Cross-entropy Loss\\
    AAE&African-American dialectal English\\
    NHB (NHW)&Non-Hispanic Black (White)\\
    FPR (FNR)&False Positive (Negative) Rate\\
    FPED (FNED)&False Positive (Negative) Equality Difference\\
    pAUCED &pinned AUC Equality Difference\\
    subAUC&Subgroup AUC\\
    BPSN (BNSP)&Background Positive (Negative) Subgroup Negative (Positive) AUC\\
    GMBAUC&Generalised Mean of the Bias AUCs\\
    pB&pinned Bias\\
    FPR (TPR)&False (True) Positive Rates\\
    NPV (PPV)&Negative (Positive) Predictive Value\\
    WEAT & Word Embedding Association Test\\
    SEAT & Sentence Encoding Association Test\\
    \hline
\end{tabular}
}
\vspace{-3mm}
\end{table}

\subsection{Sampling Bias}
\label{sec:samplingbias}
The first step towards toxicity detection is to curate data samples from across the web. Among social media content, harmful content is not a majority class. On Twitter, they cover no more than 3\% of the tweets \cite{founta}. A random sampling of data thus requires going through a considerable amount of non-toxic content to get a substantial amount of toxic samples. Hence, datasets are curated by adopting heuristics that achieve a higher density of toxicity in the samples. These heuristics can introduce spurious correlations of linguistic features with toxicity labels if not tracked, therefore termed as \emph{sampling bias}. In this section, we will give an overview of the two broad methods of sampling data from web and how to measure and mitigate the sampling bias.

\textbf{Sampling techniques and effect of text source for toxicity datasets.}
In \textit{boosted random sampling}, heuristics are applied to increase the density of harmful content in an initial random sample. This sample is usually labeled for toxicity, and the keywords or user names in these samples are used to extract more samples. Meanwhile, \textit{topic-biased sampling} \cite{Salminen2020} starts by acquiring samples based on a predetermined set of keywords (hate lexicons) and socio-political topics that are commonly known to be toxic. Dataset creators have popularly used topic-biased sampling to quickly acquire samples with a higher toxicity concentration. \citet{paper19} studied six datasets (three boosted and three topic-wise sampled) and used an abuse lexicon to mark examples as explicitly or implicitly abusive. They found the boosted randomly sampled datasets to contain a lesser concentration of overall toxicity; out of them, a higher concentration was explicit abuse.
Moreover, a higher explicit abuse ratio was seen to aid in downstream performance as \emph{verbal cues for explicitly toxic labels are easier to model.} Their work was further investigated by \citet{paper13} by conducting sampling experiments on the same underlying dataset. While broadly consistent with previous studies, they discovered that all things being equal, the text source has a more significant impact on the quality of the toxic dataset curated than the underlying sampling strategy.

\textbf{Metrics to measure sampling bias.} To capture the topical and lexical biases in the curated datasets, \citet{paper8} developed two bias metrics. These measures make use of predefined hate lexicons $w' \in W'$. While $B_1$ captures the average similarity between all topical words and the hate lexicons, $B_2$ captures how likely each topic contains a keyword. The robustness in $B_1$ scores on increasing topic count indicates the new topic words to be similar to the keywords, pointing towards a topic bias. For instance, the Waseem \& Hovy dataset (referred to as the W\&H dataset henceforth) \cite{waseemandhovy} showed robustness to topics, which can be attributed to its high topic and user biases \cite{paper19,paper16}. An Arabic dataset \cite{albadiarabicdataset2018}, which was curated based on religious keywords, gave comparable $B_1$ values for all three target attributes, suggesting its coverage of origin and gender-based topics as well. The authors further showed that these biases and variations in sampling techniques impact cross dataset generalisation. In short, one can say that models trained on a dataset sampled for a set of topics do not generalize.

\textbf{Mitigation via proxy topic-sampling.}\citet{paper14} created a dataset by automating the pre-selection of keywords to reduce sampling bias. They used Reddit to find keywords from posts with more or less equal amounts of upvotes and downvotes. Such controversial topics can be used as a proxy to collect toxic speech from OSNs \cite{graumas2019}. A small sample from the tweets filtered based on these keywords was verified as more toxic than an equal-sized sample of unfiltered tweets. However, this difference was minimal, suggesting the inability of the proposed method to create a dataset with the desired density of toxic comments. Moreover, \citet{paper14} did not validate their initial claim of a lower bias in the filtered dataset.

\textbf{Mitigation via data mixing.} Topic/author bias degrades downstream performance, the W\&H dataset \cite{waseemandhovy} is a \emph{case in point.} \citet{paper19} noticed that removing top PMI terms from the W\&H dataset for the toxic labels led to a considerable drop in classification performance, suggesting a high topic bias. The dataset also contained high author bias, with more than $70\%$ sexist tweets from only two authors and about $99\%$ racist tweets from a single author. This high author bias can be attributed to the improvements in performance when encoding the user meta-information as features \cite{qianwaseemauthorbias,mishrawaseemauthorbias}. It begs whether the classification model trained on the W\&H dataset learns to discriminate the hate styles or the hateful authors' styles. \citet{paper16} tried to correct the author bias in this dataset by limiting the number of toxic comments to $250$ per user while augmenting more toxic comments from the Davidson dataset \cite{davidsondataset}. They found a substantial increase in generalisation on the unseen Hateval dataset \cite{hatevaldataset}.  \emph{However, we note that they did not account for the topical homogeneity between the train and test datasets that could have led to an increase in performance.}

\subsection{Annotation Bias}
\label{sec:annotationbias}
As toxicity is subjective, it is challenging to train machine learning models to learn what can be potentially toxic. In order to obtain these labels, we require large-scale human-aided annotations. \citet{fortunalabelstudy} showed that toxic speech datasets use ambiguous and misleading labels to refer to similar underlying categories introducing subjectivity into the annotation process. The subjectivity and discrepancy in the labeling process can lead to \emph{annotation bias}. In this section, we will broadly discuss how the quality of data labels is impacted by the annotation guidelines and the annotator's prejudice. It should be noted that there is no direct way of mitigating annotating bias via modeling alone. Obtaining unbiased annotations requires sensitization of the annotators and the society at large. We hope this section highlights the ambiguous and challenging nature of generating standard labels for toxicity.

\textbf{The effect of guidelines on the quality of annotations.} Annotation pipelines need to consider the subjective nature of hate speech \cite{rottger-etal-2022-two}. \citet{paper18} explored the effect on the reliability of annotations before and after providing the guidelines. Both annotations (guided vs. non-guided) showed a high Pearson's correlation coefficient, indicating that they captured the same underlying construct. Similarly, the Jigsaw\footnote{\url{https://jigsaw.google.com/the-current/toxicity/}} team adopted a labeling schema that is subjective to individual interpretation \cite{paper20}. The team found that more annotators could agree on the toxicity of comments based on a more generic guideline. On the contrary, \citet{portuguesedatasetinpaper8,fortunalabelstudy} suggested using hierarchical multi-class annotation schemes.

\textbf{The effect of annotators' implicit biases on quality of annotations.}
\citet{paper17} studied the influence of an annotator's expertise on toxicity labeling. They formed two groups of annotators -- experts and amateurs and found a low agreement within these groups across all labels on the W\&H dataset. They also found that the amateurs were more likely to assign a toxic label to the sample, leading to a loss in downstream performance. The ratings by the amateurs were also closer to the W\&H dataset's crowdsourced annotations. \emph{We need to investigate the impact of the expertise and sensitivity of annotators on the final labels obtained.} 

\citet{paper22} investigated the effect of the annotator's demographics on the quality of resulting classifiers. They used the annotators' demographic information (\textit{gender}, \textit{age}, \textit{education} and \textit{native English speaker}) present in the Wulczyn personal attacks dataset \citet{wulczyndataset}. They found no differences between the two genders. Whereas, models trained on native English speakers' annotations outperformed the ones for non-native speakers in terms of $F1$ score and sensitivity. Similar differences were observed across both groups for age and education. \emph{While this investigation relied on the per-annotator demographics collected during the annotation process, this information can be tricky to collect and limited to the demographic strata considered during the data creation phase.}

Meanwhile, \citet{paper23} adopted an unsupervised method to group the behavior of annotators. Annotators and the inter-annotator agreements were converted into a graph with the annotators as nodes and their agreement scores as edge weights. Next, a community detection algorithm \cite{louvaincommunitydetection} was used to group similar annotators. They trained toxicity detection systems on annotations by groups with low inter-rater agreements, which resulted in low-performance on the test datasets labeled by other groups. Similar results were seen for a group with higher inter-rater agreement but low agreement with the other groups, indicating a bias in this group's combined annotation. However, the study did not mention obtaining multiple observations for a group, opening the question of variance in performances due to random model initialization. \emph{An exciting extension to this study would be analyzing the type of biases shown by such anomalous groups and if their ratings should be penalized or contrarily given higher weights to improve the inclusion of the demographic represented by them.}

\textbf{The need to incorporate disagreement in toxicity modeling.} In an intriguing study to account for annotation disagreements, \citet{disagreement_deconvolution} developed a disagreement deconvolution setup to estimate the lower bounds on model performance accounting for variation in human annotations. They compared the drop in performance against the oracle system for respective models trained on datasets covering subjective tasks like toxicity vs. objective tasks like image detection. Researchers observed that when disagreements are not accounted for --(i.e., taking a single label per sample), the performance drop is more for subjective tasks than objective ones. Consider a crowdsourced toxicity sample that receives $5$ annotations ($3$ non-hate, $2$ hate). When we go by majority voting, the model will consider predicting the label vector as [1,0]. Meanwhile, the probability distribution it should learn is [0.6,0.4]. Framing subjective tasks via majority labeling fails to capture the latent distribution of variable human understanding of toxicity. \emph{The study crucially points out the importance of accounting for disagreements to project the realistic performance of models when deployed in real-world toxicity detection.}

\subsection{Lexical Bias}
\label{sec:lexicalbias}
The keyword-based heuristics in data sampling, coupled with a lack of context and human predilection, can cause toxic speech classifiers to learn spurious lexical correlations. The models produce high toxicity ratings for non-toxic texts due to a disproportionately high presence of certain terms (or phrases) in the content labeled as toxic. This phenomenon is called \emph{lexical bias}. We refer to terms contributing to lexical bias as \emph{bias sensitive words} (BSW) \cite{paper21}. This section highlights mechanism to mitigate lexical bias that can be mainly achieved via (a) correcting data labels, and (b) debiasing at the modeling level.  

\textbf{Deciding on the target BSW.} 
While \citet{paper20} and \citet{paper7} relied on the manual creation of a list of identity terms, \citet{paper15} used the identity attributes annotated in the Civil comments dataset\footnote{During dataset creation, apart from the toxicity labels, samples were also annotated with the presence of lexical markers of $24$ identity attributes.} \citet{borkanmetrics2019}. \citet{paper6} used local mutual information (LMI) to find the terms highly correlated with the toxic labels. Meanwhile, \citet{paper21} developed unsupervised methods to find the set of BSW. 

\textbf{Mitigation through data correction and filtering.} \citet{paper20} and \citet{paper21} proposed new methods for mitigation of lexical bias. On the other hand, \citet{paper10} investigated methods that have been effective on other NLU debiasing tasks \cite{aflite,datamaps}.
\begin{itemize}[noitemsep,topsep=0pt]
    \item \textbf{Length sensitive upsampling}:
    \citet{paper20} found that toxic samples for an identity term are non-uniformly distributed across different lengths. Subsequently, the non-toxic label was upsampled using statements from Wikipedia articles across the length buckets. The authors also observed an improved pAUCED (pinned AUC) value on a synthetic bias evaluation set (SBET) containing samples equally distributed across the toxicity labels for each identity term. Moreover, the FPR (False Positive Rate) values across the identity terms went down without an increase in the variance of the FNR (False Negative Rate) values, indicating a reduction in false-positive bias while not affecting the false-negative bias. \emph{Note that these observations were made on SBET, which employs a static target-attribute pair list. Since this list had a different distribution than the initially-debiased data, calculations of bias based on SBET may not capture the changing dynamics/semantics when the samples are debiased}. We note the same in our case study of bias-shift (Appendix \ref{sec:casestudybiasshift}).
    
    \item \textbf{Knowledge-based generalisations:} \citet{paper21} developed mitigation techniques based on knowledge-based generalisations. They propose   lexical database generalization; replacing the BSW occurrence with an ancestor in the wordnet \cite{wordnet} hypernym-tree (e.g., \textit{black} with \textit{color}). Similar to \citet{paper20}, \citet{paper21} did not account for the newly acquired biases in the dataset post-mitigation. We present a case study related to this observation in Appendix \ref{sec:casestudybiasshift}.
    
    \item\label{it:data_filter} \textbf{Data filtering:} \citet{paper10} explored two automated data filtering approaches (AFLite \cite{aflite}, and DataMaps \cite{datamaps}) to obtain the set of training samples that will lead to better generalisability and reduce bias as a by-product. The correlation of the toxicity label in the Founta dataset \cite{founta} with BSWs reduced with both AFLite and DataMaps, corroborating a reduction in the lexical bias in the resulting filtered datasets.
\end{itemize}
 
\textbf{Mitigation through the debiased training of downstream models.} \citet{paper7,paper15} and \citet{paper10} studied the mitigation of lexical bias through model-level debiasing.
\begin{itemize}[noitemsep,topsep=0pt]
    \item \textbf{Regularising importance scores}: \citet{paper7}  reduced the weightage a model assigns to identity terms by regularising cross-entropy loss ($L_{CE}$):
    \begin{equation}
    L = L_{CE} + \alpha \sum_{w\in x\cap S}{[\phi(w)]^2}
    \end{equation}
    where $x$ is the set of input sequence words, $S$ is the set of identity terms, $\phi(w)$ is the attention score for term $w$, and $\alpha$ is the strength of regularisation. The BERT models trained on the GHC \cite{ghcdataset}and Stormfront \cite{whitesupremacystormfrontdataset} datasets with regularised loss showed a reduction in FPR on the test datasets. Using Sampling and Occlusion \cite{Jin2020Towards} as an explanation algorithm, the authors also validated a reduction of the importance score given by the regularized model to the identity terms, suggesting a decrease in lexical bias.
    
    \item \textbf{Multi-task learning}: \citet{paper15} used multi-task learning (MTL) to learn both the toxicity and identity attribute labels for the Civil comments dataset. This loss $L$ accounts for the cross-entropy loss $L_{CE_k}$ for each identity $k \in [0-9]$ in addition to the loss for the toxicity task $L_{CE}$.
    \begin{equation}
    L = \sum_{n=1}^{N}{\beta_n[\alpha L_{CE}(\hat{y}_n,y_n) + (1-\alpha)\sum_{k=1}^{9}{L_{CE_k}(\hat{y}^{k}_{n},y^{k}_{n})}]}
    \end{equation}
    where $\beta_n$ is a sample weight, given a higher value for a non-toxic example $n$ with at least one identity attribute present. Evaluation of the MTL model on the test split showed an improvement in terms of -- (i) AUC, suggesting a better overall model performance, (ii) Generalized Mean of Bias AUC (GMB AUC), suggesting better bias reduction, and (iii) Background Positive, Subgroup Negative (BPSN AUC) and Subgroup AUC (subAUC) for each identity term, suggesting a reduction in the false-positive bias while not increasing the false-negative bias. \emph{However, this mitigation method relied on the annotation of identity attributes which is not easy to extract and scale.}
    
    \item \textbf{Ensemble based debiasing}: \citet{paper10} explored the ensemble-based debiased training method on the Founta dataset. They trained an ensemble of two models, an SVM classifier and a RoBERTa-based classifier. The idea was to let the naive model learn to predict the toxic label based on the biased features, encouraging the robust model to rely on the other unrelated features. Finally, once the ensemble was trained, the naive model was discarded. The final model showed lower $FPR$ values on the BSWs containing samples from the Founta dataset, indicating better debiasing. However, it came at the cost of loss in accuracy and $F1$ score.
\end{itemize}

\begin{figure}[h]
    \centering
    \includegraphics[width=8cm]{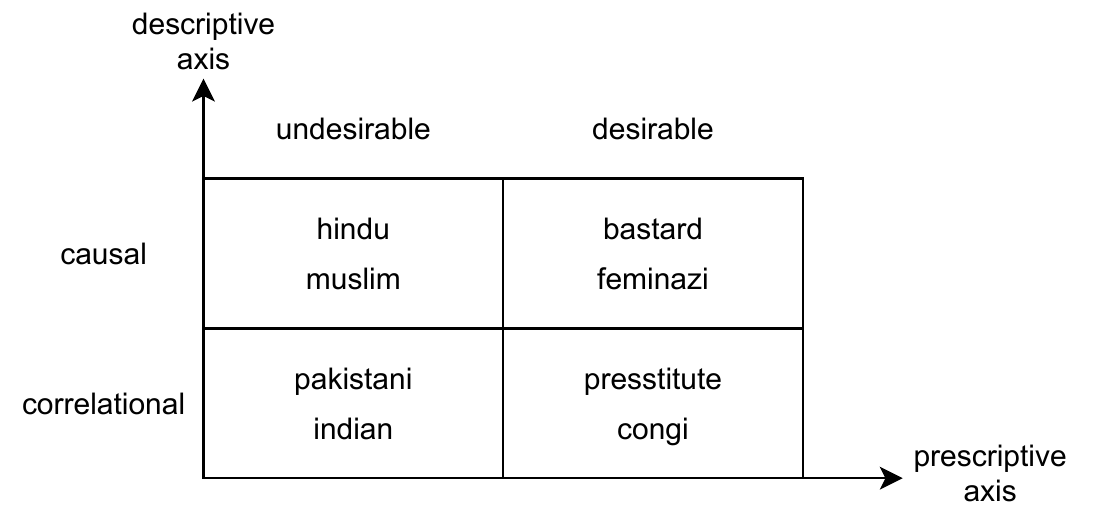}
    \caption{\citet{ghosh-etal-2021-detecting} proposed two axes: (i) descriptive and (ii) prescriptive, to segment the associations of the \emph{over-represented} terms for the model under bias evaluation. Here we consider the cross-section of transferring models trained on English toxicity datasets to toxic phrases in code-mixed Hinglish. The terms in the undesirable quadrants (left column) are those which, despite their frequent presence in toxic classes, are non-toxic/neutral and should not contribute towards toxic label prediction. Meanwhile, the terms in the desirable quadrants (right column) are offensive terms well known in English or colloquial demographic (Indian). For example, while we want the model to correctly predict abusive terms such as bastard or feminazi, or commie to capture toxic intentions, we also want the model to extend to novel lexical terms known to be offensive in the Indian context. For example, `congi' is an offensive connotation to Indian National Congress, a left-leaning political group in India.}
    \label{fig:geographicalbias}
    \vspace{-3mm}
\end{figure}

\section{Biases based on target of harm}
\textbf{Overview.} The biases introduced during data collection and annotation directly/indirectly impact the ability of toxicity models to predict toxicity against identity groups like race, gender, age, etc. Such unintended biasing of models against identity attributes manifests due to specific lexical cues present in the data that are erroneously correlated with class labels. As depicted in Figure \ref{fig:geographicalbias}, these can be linked to either the presence of \textit{non-toxic terms} in high proportion in \textit{toxicity labels} or to the presence of \textit{abusive terms} in high proportion in \textit{non-toxic labels}. For example, the co-occurrence of terms such as ``Muslims'' and ``Islam'' in comments marked as Islamophobic leads the model to correlate Muslims with toxicity spuriously. Meanwhile, the presence of phrases like ``f*** the world, be you," even though intended to be non-toxic, ends up being annotated or predicted as toxic due to the previously-learned relation of slur terms with toxicity.

Meanwhile, the focus on abusive terms or reclaimed slurs falsely causes the models to classify non-toxic content as toxic. The primary goal of mitigating biases against target groups thus aims to reduce the attention assigned to the presence/absence of specific terms and shift the focus to the overall context in the content under consideration. One way of achieving this is by providing external context as meta-data or incorporation of present world knowledge databases. In current literature, toxicity detection and debiasing employ large language models that are well-known to be biased in their understanding of identity groups. \emph{Thus, any debiasing of toxicity models is lower bounded by the debiasing of the underlying language frameworks and embeddings employed in the detection models.}

\subsection{Racial Bias}
\label{sec:racialbias}
\textbf{Racial bias in machine learning.} In their seminal work, \citet{pmlr-v81-buolamwini18a} pointed out that facial recognition systems are biased against dark-skinned faces. Similar issues were observed against Black dialects in automatic speech recognition \cite{tatman-2017-gender}, and the negative association of Black names in popular word embeddings \cite{doi:10.1126/science.aal4230}. The issue of biases in word embedding is a much-researched topic in NLP, and its impact extends to downstream tasks of toxicity detection as well. Researchers observed that content posted by Black users or content employing non-white dialects is more likely to be mislabelled as toxic \cite{paper5}. Owning to which the toxicity detection models risk further discrimination of marginalized groups. While some of the broader literature around dialect detection is important to applications like chatbots and automated speech systems, recognizing the dialect based on written content is tricky and contended. The initial work in dialect detection in text revolved around pointing out shortcomings of parsing systems that fail the grammar of diverse dialects; they have been recently adopted to study the impact of dialect in the classification of sentiment and toxicity. 

Unfortunately, most literature addressing racial bias primarily focuses on African-American dialectal English (AAE). In our discussion of racial bias in toxicity detection, we will discuss research engaged in evaluating and mitigating racial discrimination against posts attributed to Black users. Several resources have been developed to estimate a poster's race, as these attributes are only sometimes available.

\textbf{Disclaimer on dialect detection.} While some of the broader literature around dialect detection is important to applications like chatbots and automated speech systems, recognizing the dialect based on written content is tricky and contended. The initial work \cite{jacob,eisenstein-etal-2011-discovering}  in dialect detection in text revolved around understanding the changing patterns linguistics; they have been recently adopted to study the impact of dialect in the classification of sentiment, and toxicity \cite{paper1,paper10}. However, one has to note that detecting dialect can be a proxy for race and opens up the ethical implication of dialect detection and its application. The implication of the work by  \cite{blodgettlm} and  \cite{paper3} around automatic dialect detection can thus be problematic unless the context of the user's background is considered. For example, one can easily set up spam accounts on social media to imitate and defame a specific demographic region.

\begin{itemize}[noitemsep,topsep=0pt]
    \item \textbf{Blodgett LM and dataset \cite{blodgettlm} \label{sec:blogdgettlm}:} This language model was trained on tweets located in the USA and matched them with the USA census' demographic data for the four largest racial/ethnic groups (namely, those of non-Hispanic Whites (NHW), non-Hispanic Blacks (NHB), Hispanics, and Asians). Out of these, the authors used the language model to filter $1.1M$ (and $14.5M$) Black-oriented (and White-oriented) tweets by likely NHB (and NHW) users. \emph{Using Twitter as a source to train for dialects, Blodgett LM (Language Model) could very well employ the same negative word association and stereotyping when predicting dialects, as is the case with toxicity detection models. Utilizing such LMs as black-box for future dialect prediction can exponentially impact the biases against non-white speakers \cite{paper5}. Methods that use human-annotated dialect labels or do not require ground truth for measuring bias \cite{10.1145/3461702.3462557} should be favored over dialect-proxy LM-based models.}
    \item \textbf{Pietro dataset \cite{pietrodataset}\label{pietrodataset}:} It is a corpus of $5.4M$ tweets from $4132$ survey participants who reported their race/ethnicity ($3184$ NHW and $374$ NHB).
    \item \textbf{Huang dataset \cite{paper3}:} It is a multilingual dataset of tweets combining existing datasets from five languages annotated for toxicity. Four demographic attributes (including race) of the users who posted the comments were also annotated using their Twitter profiles.
\end{itemize}
In the remainder of the paper, we refer to tweets marked as NHB-related (or NHW-related) as Black-oriented (or White-oriented) tweets. We let go of common disputed terminologies like Standard-American English (SAE), Mainstream US English, etc., to not promote minority exclusion \cite{saeavoidrosa,paper10}.  \citet{paper3} noted that racial information encoded in a tweet could introduce downstream harm. They trained models on the English subset of the Huang dataset to predict the author's race solely from the tweet's text.  Contrary to the findings of \citet{paper5}, they found that words like \textit{n*gga} and \textit{b*tch} were more significantly related to the NHW class, suggesting a derogatory use of these terms. \emph{It is an exciting finding as the Huang dataset is one of the few works not relying on the Blodgett LM for proxy of the user's race.}

\textbf{Racial information can lead to annotation bias.}
Black-oriented tweets often get mislabeled as toxic. \citet{paper1} showed a positive correlation between $p(NHB|tweet)$ and the toxic labels for the Davidson and Founta100k. Similarly, \citet{paper2} applied structural topic modeling (STM \cite{stm}) on the Founta100k dataset and identified a latent topic containing terms prevalent in AAE (e.g., *ss, n*gga, etc.). to be more likely to be flagged as toxic. 

\textbf{Addressing the biased annotations.} \citet{paper1} and \citet{paper10} explored the mitigation of racial bias by addressing annotation bias.
\begin{itemize}[noitemsep,topsep=0pt]
    \item \textbf{Annotator priming:} \citet{paper1} discovered that annotators were more likely to label a sample as ``non-toxic" when primed with the dialect or race of the author. However, it must be noted that annotator priming is a tricky task. Sometimes, it can nurse the annotators' implicit human biases instead of suppressing them.
    \item \textbf{Dialect aware label correction\footnote{\citet{paper10} cautioned against a real-world application of this method due to limitations of GPT-3.}:} \citet{paper10} utilised the few-shot capabilities of the GPT-3 \cite{gpt3} by supplying a few seed examples \cite{aaelanguageuse} for Black-oriented to White-oriented dialect conversion. They used the Founta100k dataset and the Blodgett LM to find the \textit{two *-oriented tweet divisions}. A Black-oriented sample is marked as non-toxic if its White counterpart is labeled non-toxic. The language model trained on the relabeled dataset led to a lower false positive rate on the Black-oriented test samples.    
\end{itemize}

\textbf{Diagnosing bias in downstream models.}
\citet{paper1} showed that a downstream model trained on either Davidson or Founta100k dataset produced: (i) higher FPR values for the black-oriented comments, and (ii) higher FNR values for the White-oriented comments of the test-split. Higher positive rates were also observed for Black-oriented tweets from the Blodgett and Pietro datasets. Similarly, \citet{paper3} observed higher FPED and FNED values by downstream models trained and tested across all languages of the Huang dataset. Meanwhile, \citet{paper5} calculated the proportion of Black-oriented ($p_{black}$) and White-oriented ($p_{white}$) tweets from the Blodgett LM predicted as toxic by the models trained on respective train set. A $t$-test on five popular Twitter toxic speech datasets revealed a significant disparity against the Black-oriented tweets. This disparity reduced (yet persisted) when the $t$-test was repeated conditioning on the presence of identity terms. The authors noted that while this prevailing disparity could have been due to other terms not conditioned upon, it was also possibly the result of the model correlated other subtler details of AAE with the toxicity.

\textbf{Mitigation through the debiased training of downstream models.} \citet{paper6,paper11,paper10} explored mitigation of racial bias through model debiasing.
\begin{itemize}[noitemsep,topsep=0pt]
\item \textbf{Comment re-weighting in loss:} Following  \cite{factverificationmodels}, \citet{paper6} defined a bias score $s^{c}_{j}$ for each label class $c$ and  bigram term $t_j$ as:
    \begin{equation}
    s^{c}_{j} = \frac{\sum_{i = 1}^{n}{I_{t_j\in x_i}I_{y_i = c}(1+\alpha_i)}}{\sum_{i=1}^{n}{(1+\alpha_i)}}
    \end{equation}
    where $\alpha_i$ is the weight of the $i^{th}$ sample, and $I$ is the indicator function. These sample weights $\alpha$ were learned by solving the following optimization:
    \begin{equation}
        min(\sum_{j=1}^{|V|}{max_c(s^{c}_{j}) + \lambda||\alpha||_2})
    \end{equation}
    where $V$ is the set of bigrams in the train data, and $\lambda$ is a hyperparameter. Though the debiasing was applied to the complete vocabulary, mitigation only in racial bias was evaluated. Reduction in bias against the Black-oriented tweets across all toxicity labels was observed for models trained on the W\&H and Davidson datasets. 

\item \textbf{Adversarial training:} \citet{paper11} used an adversarial training \cite{kumar2019demotinglatentconfounds} on the Founta100k dataset. One classification head ($C$) predicted the toxicity label, while the adversary head ($D$) predicted the protected attribute $p(NHB|tweet)$.
This method was shown to be effective when lexical cues between the two attributes are closely related (e.g., terms like \textit{n*gga} and \textit{b*tch} correlate with both the AAE dialect and toxicity labels \cite{paper5}). The model showed a reduction in FPR and an increase in macro-$F1$ scores. 
    
\item \textbf{Ensemble based debiasing:} Similar to lexical bias (see Section~\ref{sec:lexicalbias}), \citet{paper10} utilised LearnedMixIn for mitigating racial bias on the Founta dataset. They used the four $p(*|tweet)$ predictions from the Blodgett LM as features for the naive model. A drop of FPR on the Black-oriented comments was observed, along with a drop in macro-F1. \emph{However, an insignificant drop in disparity against the Black-oriented tweets were observed when tested on the Pietro dataset, suggesting the inability of mitigation methods on out-of-domain data.}
\end{itemize}
    
\subsection{Gender Bias}
\textbf{A note on gender in NLP.} \citet{bolukbasi} pointed out the gendered bias in word embeddings considers a pair of masculine or feminine attributes and job occupations. Since then, multiple studies have evaluated gender bias \cite{sun-etal-2019-mitigating, van-der-wal-etal-2022-birth} pinned on the binary gender identity. Only a few studies have explored this phenomenon from a non-binary perspective \cite{zhang-etal-2020-demographics}. Talking specifically of toxicity detection systems, Jigsaw\footnote{\url{https://www.kaggle.com/c/jigsaw-unintended-bias-in-toxicity-classification/}} is probably the only dataset that considers more inclusive and fine-grained labels for gender and sexual orientation as the target of toxicity. Unfortunately, the study of non-binary gender bias is missing from the toxicity literature.
\label{sec:genderbias}

\textbf{Evaluating gender bias.}
\citet{paper17} showed that a model trained on the W\&H dataset led to a performance gain by including the user's gender as a feature \cite{saplexica}. It stems from the fact that abusive gendered terms like `b*t**', `wh***' are more likely to be observed in samples labelled as toxic.
\citet{paper9} compared the FPED and FNED values using an SBET \cite{genderbiasinnlpsurvey} model trained on the W\&H and Founta datasets. They found that using pre-trained embeddings improved toxic speech detection performance at the cost of more significant equality difference scores. The authors also observed higher bias values for the W\&H dataset than the Founta dataset, attributable to Founta's bias-aware design (Section~\ref{sec:samplingbias}).
 
\textbf{Mitigation techniques.} Based on their observations, \citet{paper12}, and \citet{paper9} further proposed mitigation techniques for gender bias:
\begin{itemize}[noitemsep,topsep=0pt]
\item \textbf{Debiased embeddings:} \citet{paper9} applied three strategies for gender bias mitigation on the W\&H  dataset: (i) using debiased-word2vec embeddings \cite{bolukbasi}, (ii) gender-swap data augmentation (using gender pairs identified by \cite{zhao-etal-2018-gender}), and (iii) transfer learning by first fine-tuning on a less biased large-scale dataset (Founta \citet{founta}).  The best debiasing performance was achieved by combining debiased embeddings and gender-swapping, reducing FPED and FNED. Meanwhile, the transfer learning approach led to the highest AUC loss. \emph{We attribute this to incompatible labels between the source and the target datasets (abusive vs. sexism).}
\item \textbf{Length sensitive upsampling:} \citet{paper12} identified $12$ common female identity terms in the Evalita2018 \cite{evalitadataset} dataset. Next, they upsampled tweets from the W\&H dataset to balance the count of occurrences of each identity term across the toxic and non-toxic classes of the training dataset. Upsampling while considering the tweet length ranges (similar to \citet{paper20}) led to the best gender debiasing. This debiasing, however, came at the cost of a small AUC drop on the test split. \emph{The W\&H dataset is known to contain a host of biases which can affect the resulting upsampled dataset.} \cite{paper19}
\end{itemize}

\subsection{Intersectional bias}
\label{sec:intersection_bias}
\citet{kim2020intersectional} study the intersection of racial and gender bias for AAE. They used the Founta100k dataset and labeled race using the Blodgett LM (c.f. Section \ref{sec:racialbias}). They also marked the dataset with gender and party (political inclination). The party information is used as a control variable. While tweets belonging to the Black community are more likely to be classified as abusive, Black males are significantly more likely to be classified as hateful. In toxic speech detection, it is essential to evaluate such intersectional biases more carefully.

\subsection{Cross-geographic bias}
\citet{ghosh-etal-2021-detecting} observed that most literature in toxicity detection focuses on the English language; this concentrated attention towards a few geographies creates a \emph{knowledge gap} and can lead to lexical bias. As a concrete example; they showed that Alphabet's Perspective API\footnote{\url{https://perspectiveapi.com/}} gives lower toxicity scores to terms that are considered toxic in Indian context (for example, \textit{presstitute}, a slang combining the terms \textit{press} and \textit{prostitute}) leading to false negative predictions. Conversely, terms such as \textit{muslim} and \textit{hindu} were observed to generate higher toxicity scores even in a non-toxic context, leading to false positives. A set of reproduced terms from their study is provided in Figure \ref{fig:geographicalbias}. \citet{ghosh-etal-2021-detecting} then proposed a two-step weakly-supervised method to detect lexical bias for cross-geocultural harmful content. They carried out this analysis using unlabeled tweets collected from seven countries. In the first step, they identified a set of terms $T$ that were statistically overrepresented in the tweets from a country. Their study separated $T$ along two axes concerning the model under investigation: (i) descriptive axis: correlational vs. causal associations, and (ii) prescriptive axes: desirable vs. undesirable associations. Correlational associations refer to the terms not invoking toxicity in the model, while causal refer to the terms causing higher toxicity predictions. Similarly, desirable associations refer to the terms for which higher toxicity ratings are desirable and vice-versa for undesirable. In the second step, to find the correlational associations, they built $33$ hand-curated template sentences with varying degrees of toxicity (for example, ``You are a <person>'', ``I dislike <person>''). They replaced the template with toxic entities in $T$ and recorded the change in toxicity score. Any term in $T$ causing the toxicity score to increase while it was not desired to is an instance of spurious correlation. They then employed various mitigation techniques like deletion, substitution, and balancing and found that these methods did not significantly reduce bias.

\subsection{Political bias}
\label{sec:lastbiassection}
Recently, \citet{paper24} studied the effect of political bias on hate speech detection. Tweets were collected from political leaders who represented the left-winged, right-winged, and neutral ideologies, with each tweet heuristically marked as non-toxic. Finally, these three politically inclined corpora were used to replace non-toxic comments from an existing corpus \cite{germeval2018,germeval2019}, resulting in politically-biased datasets. It was observed that the right-wing biased dataset produced statistically lower $F1$ lower than the other two biased datasets—\emph{the work did not explore introducing political bias for the toxic label. We find this assumption unconvincing that politically-based tweets were treated as only non-toxic.}

\section{Major takeaways: A summary}
\label{sec:takeways}
This section highlights the key takeaways from surveying existing debiasing methods. The summary table of various debiasing methods for the respective bias is provided in Table \ref{tab:takeway}. We also compare physical systems and bias as physical entities. This analogy highlights the drawbacks of existing debiasing methods and motivates a continuing integrated system for mitigating biases.

\textbf{Inspiration from physical systems.} We extend the interpretation that ``\textit{energy can neither be created nor destroyed but only transformed from one form to another}'' to bias as a physical entity. One can say that as long as humans are bombarded with more information than they can process, some form of cognitive biases will keep existing and evolving as per the social structures of the zeitgeist. It also means that despite our best efforts, our interactions (online or offline) will be riddled with biases and stereotyping -- the same manifests as biases in downstream tasks of NLP. Just like the ``\textit{system at rest remains at rest unless acted upon by an external force,}'' to consciously and effectively overcome biases, we need to supply our NLP pipelines with external impetus in the form of debiasing evaluations and criteria. However, one-time mitigation and evaluation of bias, as tested in the controlled setting by all the techniques discussed in this survey, will unfortunately not be an accurate (or even valid) indicator of bias reduction as the system evolves. The principle that the ``\textit{entropy of a closed system never decreases,}'' renders our one-time bias mitigation efforts moot. Translating entropy to the close-minded biases we harbor, our biases, once seeded, will not automatically disappear. Instead, bias, as an entity, is expected to reinforce and strengthened with time. i.e., years of systematic discrimination and marginalization leading to increased instances of hate speech (offline and online) and hate crime. Thus, we need a continuously probing setup (a regular supply of external energy) to keep the biases in the NLP pipeline in check. We can only ensure their mitigation in an ever-evolving system of human interactions by evaluating biases at every step of the workflow, and analyzing it at regular intervals.

\begin{itemize}[noitemsep,topsep=0pt]
    \item \textbf{Categories of Bias:} As we observe in Figure \ref{fig:datatransformations}, the bias categories based on the \textit{sources} of harm are not necessarily exclusive. Lexical bias can be a consequence of sampling or annotation bias or both. Similarly, any category of bias based on downstream harms can be an artifact of all three categories based on the source of harm.
    
    \item \textbf{Sampling Bias:} These observations collectively suggest that the text source and the topics used for sampling have a greater influence on the bias characteristics of the dataset, than the sampling strategy. Additionally, OSNs have varying tolerance towards toxic speech \cite{Munn2020}. The difference in platform specific policies can further affect the bias characteristics of the sampled dataset.
    
    \item \textbf{Lexical Bias:} There should be a balance of sensitive words and phrases like explicit abuse, identity terms (especially that of minority groups), and topic words across the labels to not fabricate any spurious lexical correlations with the labels. These correlations can disrate the model's ability to capture the context of such terms. A common point of failure of models trained on such data is the conflation of identity disclosures (e.g., `I am gay')  with identity attacks (e.g., `I hate all gays'), further promoting disparity amongst social groups.
    
    \item \textbf{Annotation Bias:} The current literature for annotating harmful content is spread on the spectrum from strict to loosely defined guidelines. Consequently, we observe that the annotation agreements also vary, with no fixed benchmark for toxic content labeling. The observations and results from the above sections highlight the importance of examining the sensitivity of the annotators towards socio-cultural factors such as dialects and race. In order to effectively study annotators' behavior across attributes, it is essential to have a fair number of participants.  
    
    \item \textbf{Racial Bias:} Every mitigation strategy suggested to reduce racial bias in existing models assumed that both Black-oriented and White-oriented samples follow the same conditional probability $p(Y|X)$. This assumption is flawed because using specific terms is socially acceptable for a Black person while unacceptable for someone else. Though the resources from \cite{blodgettlm,pietrodataset,paper3} have aided the investigation of racial bias in toxic speech, none of them have access to the ground-truth labels for the dialect identities. For example, it is possible that the Blodgett LM spuriously correlates the presence of terms like `\textit{n*gga}' and `\textit{b*tch}' with the tweet being Black-oriented \cite{paper5}. Such predictions can magnify the racial bias during toxicity detection. Moreover, the assumption that only certain social groups employ specific terms is flawed in today's era. While on the one hand, it leads to broader adoption and acceptance of cultural terms, on the other hand, such terms are employed by extremist groups and trolls for misappropriation and mocking other cultures. Without knowing the background and intent of the online users, it is difficult to pinpoint what they wish to achieve when adopting non-native terms and dialects. The lack of awareness about dialects and their cultural importance among NLP researchers can itself be a source of bias.
\end{itemize}

{\color{blue}
\begin{table}[!t]
\caption{A summary of debiasing methods employed for different types of unintended biases as discussed in this survey.}
\label{tab:takeway}
\scalebox{0.8}{
  \begin{tabular}{ | l | p{0.9\textwidth} |} 
    \hline
    \textbf{Bias} & \textbf{Debiasing Method}\\
    \hline
    Sampling & Topic proximity \cite{van-der-wal-etal-2022-birth}, Data mixing (augmentation) \cite{paper16}\\
    Annotation & Disagreement deconvolution \cite{disagreement_deconvolution} \\
    Lexical & Length sensitive sampling \cite{paper20}, Knowledge-based term generalization \cite{paper21}, Data filtering \cite{paper10}, Regularization \cite{paper7}, Multi-task learning \cite{paper15}, Ensemble modeling \cite{paper10} \\
    Racial & Annotation priming \cite{paper1}, Label correction \cite{paper10}, Regularization \cite{factverificationmodels}, Adversarial training \cite{paper11}, Ensemble modeling \cite{paper10} \\
    Gender & Length sensitive sampling \cite{paper12}, Debiased LM \cite{bolukbasi}, Gender Swapping \cite{zhao-etal-2018-gender}, Transfer learning \cite{paper9} \\
    \hline
\end{tabular}
}
\vspace{-4mm}
\end{table}}

\section{Direction for future research}
This section discusses the common challenges across bias mitigation in toxic speech and their possible solutions. Some of these challenges point to the general area of toxicity detection modeling and has been touched upon in other survey on toxic speech \cite{chakraborty2022nipping,generalisable-hs-survey}. 
\begin{itemize}[noitemsep,topsep=0pt]
    \item \textbf{Cognizance towards side-effects.}
    Owing to resource constraints, the study of bias and its mitigation has focused on reducing only one bias at a time. While bias mitigation is vital for toxic speech detection, it is important to acknowledge its ability to introduce newer biases in the pipeline. As stated earlier, the taxonomy of bias is overlapping in nature. For example, lexical bias can be a source of racial and gendered harm. Meanwhile, the lexical bias could be introduced due to spurious data collection and annotation biases. We observe that researchers often failed to acknowledge this critical aspect while evaluating their proposed mitigation methods (Appendix \ref{sec:casestudybiasshift}). Initial work in the direction of intersectional bias has been led by \citet{kim2020intersectional} who analysed the combined impact of gender and race (Section \ref{sec:intersection_bias}). However, the analysis and evaluation of the inter play of various biases on toxicity detection remains an open question.
    
    \item \textbf{Data collection and annotation.} As observed and discussed in  Section \ref{sec:samplingbias}, the source and topic of content can have an overarching impact on the characteristics of the dataset curated. While random sampling is closer in characteristics to the real-world distribution, they are highly skewed towards non-hate, which makes collecting toxic comments hard. Meanwhile, priming for specific topics, hashtags, or users to increase the toxic content introduces unintended biases into the dataset and the modelling pipeline. Recently \citet{rahman2021an} proposed an information retrieval (IR) based approach to collecting hate speech from Twitter. Their IR inspired method increased the coverage of hate compared to existing datasets. Such cross-domain methods can help increase the relevance of the content that should be filtered for labelling. Keeping the biases in check can lead to better generalisation. Meanwhile, proposing and standardizing code books for data annotation of toxicity labels can help reduce the variability in labels employed by different datasets. A case study is the large-scale harassment annotation code book \cite{golbeck_harrase} that helped label $35k$ instances of various categories of online harassment, including hate speech. When researchers want to propose a new toxicity class, a clear distinction and relation to existing labels should be made. Another exciting area of research can be employing weakly supervised or unsupervised methods of data augmentation \cite{Sarwar_Murdock_2022} and domain adaption \cite{ludwig-etal-2022-improving} to reduce the need for annotations and improve generalizability of toxicity detection models. 
    
    \textbf{Gap in social and computational understanding of toxicity.} It is also essential to highlight the gap between the social and psychological studies of harassment faced by vulnerable groups and its computation analysis by technical researchers and industrialists. Currently, the two systems operate in silos with little interaction between them. Such practices lead to uninformed people developing toxicity detection pipelines without a nuanced understanding of the social issues. For example, without being cognizant of the social/mental impact of objectifying women, it can be tricky to crawl public forums that capture toxicity against gendered body imaging. Therefore, we need to closely incorporate and synergize social issues, and their corresponding data sampling and annotation.
    
    \item \textbf{Gender is not just binary.} Existing literature in the area of gender debiasing in NLP as well as in toxicity detection has evaluated gender as binary (male vs female). In their recent work, \citet{dev-etal-2021-harms}  provided a general overview of how non-binary individuals are at risk of erasure and misgendering at the hands of existing language models. These harms trickle down to the task of toxicity detection as well, and unfortunately, its full extent has not been studied yet. As observed in the existing analysis of annotation and lexical biases, annotators' lack of awareness around gender fluidity can lead to inconsistent labels. More so, the toxicity models predict both "I am a homosexual" and "I hate homosexuals" as toxic due to the presence of the word "homosexual", which has been historically used to detest the LGBTQ+ community and has only recently been reclaimed.   
    
    \item \textbf{Language is not static.} Extending from the previous point, the case of word reclamation is a part of the discussion around the use of static hate lexicons, offensive dictionaries and static knowledge graphs, which cannot account for the evolving language and the evolving social-cultural aspects \cite{Steels2016}. Recently, \citet{qian-etal-2021-lifelong} proposed a prototypical learned model for hate speech classification that aims to capture the evolving hateful content as it develops. Such models need to be extended to debiasing methods as well. 
    
    \item \textbf{Proactive bias mitigation.} \citet{paper10}  showed that instead of targeting existing bias-ridden datasets, downstream harms could be better mitigated by incorporating mitigation techniques starting in the early stages of the learning pipeline, such as data sampling and annotation. While the mitigation methods proposed for annotation and sampling biases are tied to their respective steps in the machine learning pipeline, the methods for other categories are distributed throughout the pipeline. Additionally, most of the surveyed papers demonstrated bias mitigation as a single-step solution \cite{paper20,paper6}. However, it is essential to be ``bias-aware" throughout the learning pipeline \cite{paper9}. A good reference for this can be \cite{frameworksuresh}, which formalised the complete pipeline as a sequence of data transformations and defined the potential sources of harm.
    
    \item \textbf{Out of domain evaluation.} \citet{paper10} showed that mitigation techniques displaying encouraging results on in-domain samples failed to reduce disparity when tested on out-of-domain datasets. This is an alarming finding as most mitigation techniques are tested on in-domain data, putting their generalisability in question. This observation entails introducing diverse benchmarks to standardise existing and future work. For example a benchmark \cite{cvbiasbenchmark} was recently developed to systematically compare gender-bias mitigation techniques in visual recognition models.
    
    \item \textbf{Collecting user feedback.} The Jigsaw team utilized user feedback as a key source of bias mitigation\footnote{\url{https://medium.com/jigsaw/unintended-bias-and-names-of-frequently-targeted-groups-8e0b81f80a23}} for its Perspective API. Setting up of a feedback infrastructure, wherever possible, allows a collection of data that better represents the target population. 
    
    \item \textbf{World beyond English text.} In majority of literature around toxicity detection and debiasing, we mostly consider English datasets. English being the is the most widely available language on the Internet, has most accurate preprocessing tools. This has created a knowledge gap when applying toxicity systems for other languages. Especially since what can be considered toxic in English speaking geographies may not be considered toxic in other geographies. The initial study in cross-cultural bias is being led by the work of \citet{ghosh-etal-2021-detecting}. However, the extensive study of toxicity bias in non-English and code-mixed settings remain non-existent. Additionally, the workaround flagging harmful content has focused majorly on text-based features as they are easier to collect. Meanwhile, the usage of memes and videos (short clips and long ones) spreading toxic and harmful content has been gaining momentum \cite{pramanick-etal-2021-detecting,pramanick-etal-2021-momenta-multimodal,kiela2020hateful}. We need to study the impact of bias in multi-modal content as well. Do we look at the textual and image-based biases separately, or build a unifying mechanism that can capture overall-modality bias remains an exciting and open area of research. 
\end{itemize}

\section{Conclusion}
\label{sec:conclusion}
While a reasonable amount of work has been dedicated to unintended bias in toxic speech detection, we observe that its mitigation needs further exploration. Due to the multi-faceted and dynamic nature of online toxicity and its unintended biases, conducting a systematic study helps better understand the scope of the bias mitigation strategies. To our knowledge, no survey approaches this subject, focusing on the proposed methods. Therefore, we filled the gap through this survey. We developed a taxonomy of bias based on their source and target of harm. This categorization enables us to effectively discuss these methods and their drawbacks, challenges, and future work directions. We also draw attention to the need to handle more psychographic biases. For example, toxic speech based on political interests is a known issue on OSNs. However, the study of discrimination based on political leanings in harmful speech detection is still an under explored avenue \cite{paper24}. We also conducted a case study to introduce the concept of bias shift due to knowledge-based bias mitigation methods. While certainly not exhaustive, we called attention to a list of common challenges and pitfalls of bias-handling methods for toxic speech detection.

\section*{Acknowledgement}
We would like to thank the support of Prime Minister Doctoral Fellowship (SERB India), Ramanujan Fellowship (SERB, India), and the Wipro Research Grant. 
\bibliographystyle{ACM-Reference-Format}
\bibliography{survey_arxiv}
\appendix
\section{Evaluation Metrics}
\label{sec:evaluationmetrics}

\begin{figure}[h]
    \centering
    \includegraphics[width=0.8\columnwidth]{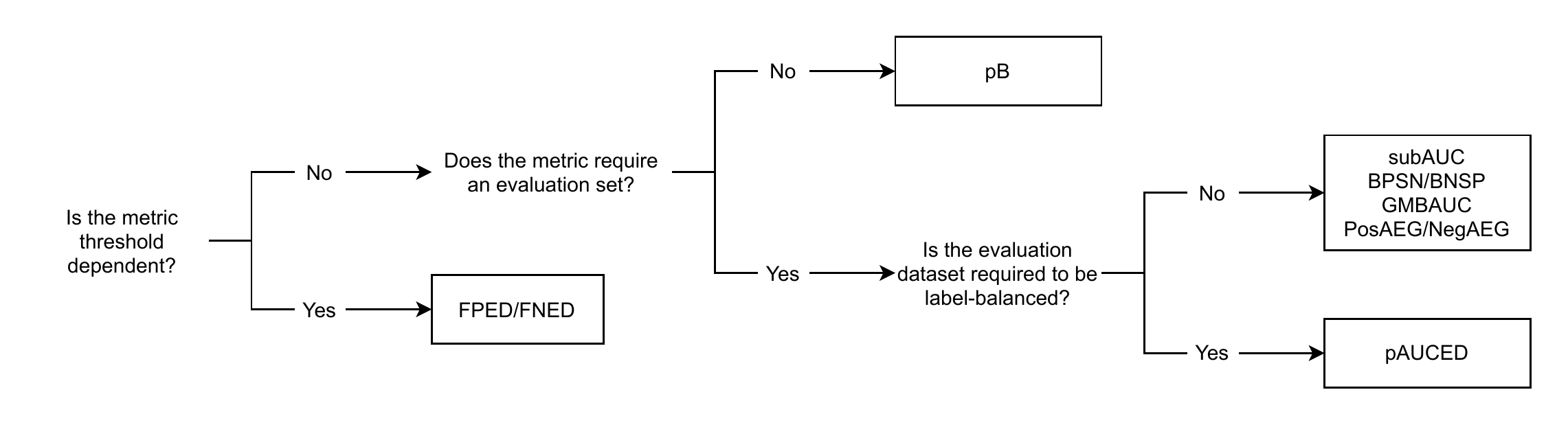}
    \caption{A taxonomy of bias evaluation metrics popular in toxic speech detection.}
    \label{fig:metricstaxonomy}
    \vspace{-3mm}
\end{figure}

\subsection{Background on bias evaluation}
This section presents about two popular bias evaluation metrics outside the task of classification. They enable us to draw motivation for the development of classification specific metrics.

\textbf{Bias evaluation based on psychological tests:}
Back in 1988 \citet{Greenwald1998} developed the implicit association test\footnote{\url{https://implicit.harvard.edu/implicit/}} (IAT) as a measure to capture the subconscious biases in humans. It is based on observing the difference in reaction times and accuracy in categorising two words settings. The first setting relates male with science and female with liberal arts. The second setting reverses the relations. In both settings, the test subjects are tasked to categorise a word towards either of these relations. It was observed that most test subjects could categorise words faster and accurately in the first relation (male = science) compared to the second (female = science). \emph{Note, the reliability of IAT is questionable \cite{Schimmack2021}, and it can at best be considered only a weak indicator of bias. \footnote{\url{https://www.apa.org/monitor/2008/07-08/psychometric}}.} 

\citet{caliskan} extended IAT to develop the Word Embedding Association Test (WEAT). It measures social biases through difference in association strengths in non-contextual word embeddings (e.g., word2vec \cite{w2v}, GloVe \cite{glove}, etc.). It should be noted that the effectiveness of WEAT as a metric heavily depends on the terms used to seed the algorithm. Later, \citet{may-etal-2019-measuring} extended WEAT for sentence embeddings, naming it the Sentence Encoding Association Test (SEAT). SEAT used templates (``This is a[n] <word>'') to insert individual words into the slot `<word>'. These templates were designed to convey minimal focus towards the context of the inserted words, helping measure the associations a sentence encoder makes with these inserted terms. Next, it applied WEAT to these synthetic sentences' embeddings instead of word embeddings. In other words, SEAT is a generalisation of WEAT on multi-word sequences.

\textbf{Bias evaluation based on finding bias subspace:}
\citet{bolukbasi} defined the gender bias of a word as the correlation between the projection of the word embedding onto the gender sub-space, and the manually annotated bias rating of the word \cite{genderbiasinnlpsurvey}. This gender subspace itself was found by applying principal component analysis (PCA) on a set of vectors obtained by taking a difference of words that vary only in the gender context (e.g., $\vec{\text{king}} - \vec{\text{queen}}$) and retaining the top component(s). \citet{sentdebias} extended this idea to contextual sentence embeddings. Unlike \citet{bolukbasi}, to calculate the gender sub-space, they used a large number of sentence pairs, with each sentence in a pair varying just in the gender-specific word. Both the bias evaluation categories described are yet to find a direct utility in the field of bias evaluation in toxic speech classification. One rudimentary exploration for the toxic speech classification task can be the analysis of social biases in fine-tuned word embeddings or sentence encoders after training them for the downstream task dataset.

\subsection{Unintended bias metrics}
\label{sec:unintendedbiasmetrics}
This subsection discusses metrics developed especially for bias evaluation of classification models. As described in Figure \ref{fig:metricstaxonomy}, we create a taxonomy of bias evaluation metrics standard in toxic speech detection literature. We list the metric abbreviations along with other commonly used abbreviations in the paper in Table \ref{tab:termsused}. In the following definitions, we assume $T$ as a set of terms $t_1, t_2, ..., t_{|T|}$. The metrics defined for a term $t$ can also be calculated for a designated group $g$ among a host of vulnerable groups $G$. 

For a test dataset $D$, the following definitions apply: (i) let $D_t^+$ be the positive (toxic) examples containing  term $t$, (ii)let $D_t^-$ be the negative (non-toxic) examples containg term $t$, (iii) let $D_{\setminus t}^+$ be the positive (toxic) examples not containing the term $t$, and (iv) $D_{\setminus t}^-$ be the negative (non-toxic) examples not containing the term $t$.

\begin{itemize}
\item \textbf{Error rate equality differences ($\mathbf{FPED}$, $\mathbf{FNED}$):}
\citet{paper20} introduced the metrics False Positive Equality Difference (FPED), False Negative Equality Difference (FNED), and pinned AUC Equality Difference (pAUCED). False Positive Equality Difference (FPED) and False Negative Equality Difference (FNED) quantify a variation of equality of opportunity. Mathematically, the authors calculated these variations of term-wise error rates ($FPR_t$ and $FNR_t$) around the error rates ($FPR$ and $FNR$) of the complete evaluation set:
\begin{equation}
FPED_T = \sum_{t\in T}{|FPR - FPR_t|}
\end{equation}
\begin{equation}
FNED_T = \sum_{t\in T}{|FNR - FNR_t|}
\end{equation}
Note that in an ideal case, $FPED_T = FNED_T = 0$.
Both FPED and FNED are threshold dependent and require a classifier that produces binary labels. However, many models produce probability distributions. \citet{paper20} and \citet{borkanmetrics2019} applied multiple threshold agnostic metrics for such scenarios.

\textbf{Usage:} In general, FNR and FPR capture the upper bound on misclassification that a model can currently achieve. When employing $FNR_t$ and $FPR_t$ on different identity terms, if the model is biased towards the terms, there will be higher variance in  $FNR_t$ and $FPR_t$, leading to higher FPED and FPED.

\item \textbf{Sub-group AUC ($\mathbf{subAUC_t}$):}  It calculates AUC on $(D_t^+\cup D_t^-)$, i.e., the examples containing the identity term $t$.

\textbf{Usage:} $subAUC_t$ based on specific terms measures the model's ability to correctly predict the toxic and non-toxic classes based on $t$. A lower value means the model is not able to distinguish the toxic samples from the non-toxic ones containing  term $t$. A higher value, on the other hand, does not mean the model is debiased. It can still give non-toxic mentions of term $t$ a high toxicity score, but as long as it gives a toxic mention of $t$ a relatively higher toxicity score than its non-toxic counter part, $subAUC_t$ will be high.

\item \textbf{Pinned AUC ($\mathbf{pAUCED}$)}: Area under the receiver operating characteristic curve (AUC-ROC or AUC) on the complete evaluation set can be insufficient at diagnosing bias, as a low AUC does not help in identifying the bias-ridden terms in $T$. Similar limitation has been noticed for $subAUC_t$. To overcome these limitations \citet{paper20} developed pinned AUC ($pAUCED_t$) for a term $t$ or a subgrroup $g$, which is the AUC measure on an auxiliary dataset $pD_t$ such that $pD_t = s(D_t) \cup s(D)$ and $|s(D_t)| = |s(D)|$, where $D_t$ is the set of comments containing  term $t$ in the evaluation set, $D$ is the complete evaluation set, and $s(.)$ is a sampling function. Here the auxiliary data allows examples from $s(D)$ to be selected at random while ``pinning" down the underlying distribution for the samples containing  term $t$, i.e., the distribution of $t$ in $pD_t$ should resemble the overall distribution of $t$ in $D$ as closely as possible. In layman's terms, one can also assume the "pinned" samples to be the "representative" samples capturing the overall distribution of the control variable under consideration. In the toxicity study, these control variables are the identity terms $t$ or the vulnerable group $g$. However, \citet{pauc-limitations} showed that the ability of $pAUCED$  to reveal bias is highly dependent on the distribution of labels between the identity terms. \citet{paper20} avoided this drawback by generating a synthetic bias evaluation set (SBET) with balanced label distribution. However, generating a SBET can be a tedious task \cite{paper21}.

\textbf{Usage:} When we have multiple subgroups $g \in G$, we can employ pinned AUC to quantify and rank the level of unintended bias of the model w.r.t various subgroups. The group with lowest pinned AUC is carrying the highest bias.

\item \textbf{Background Positive Subgroup Negative AUC ($\mathbf{BPSN}$), and Background Negative Subgroup Positive AUC ($\mathbf{BNSP}$)}: $BPSN$
calculates AUC on test set where non-toxic samples contain the bias term $t$, while toxic samples that do not, i.e $(D_t^-\cup D_{\setminus t}^+)$. Meanwhile, in BNSP AUC, we do the reverse, selecting the toxic samples that mention a term $t$ and non-toxic ones that do not, i.e $(D_t^+\cup D_{\setminus t}^-)$. The former leads to reduction in FPR, while latter leads to reducing in FNR.

\textbf{Usage:} A lower BPSN means the model confuses non-toxic mentions of the identity term ($t$) with toxicity examples that do not contain the term. In other words, the model assumes the non-toxic samples containing  $t$ to be close to the toxicity class in general. For example, when the term ``nigga" is used in a post, ``I feel for you, my nigga," the model focuses on the high correlation between "nigga" and toxicity instead of the neutral context. 

On the other hand, a lower BNSP means the model confuses toxic samples that contain $t$ with non-toxic samples that do not contain the term. Here, even the toxic usage of $t$ is assumed to be closer to the non-toxicity class. For example, the sentence ``Whites are superior to the rest" may not receive high toxicity as the identity term ``White" usually carries low negative association without context. One can blame sampling and annotation (Section \ref{sec:samplingbias} and \ref{sec:annotationbias}) biases for this.

\item \textbf{Generalised Means of Bias AUC ($\mathbf{GMBAUC}$)}: In order to obtain a mean bias score across different bias AUC metrics, Jigsaw\footnote{\url{https://www.kaggle.com/c/jigsaw-unintended-bias-in-toxicity-classification/}} introduced a generalised mean of the bias AUC (or GMBAUC) as:
\begin{equation}
GMBAUC_{p,T} = (0.25 \times AUC) + \sum_{a=1}^{3}{0.25 \times \Big(\frac{1}{|T|}  \sum_{t=1}^{|T|}{m_{a,t}^p}\Big)^\frac{1}{p}}
\end{equation} where AUC is the overall ROC AUC score, $m_{a,t}$ is the $a^{th}$ AUC-based metric calculated for term $t$, and $p$ is the power of the mean function, Inclusion of overall AUC helps in capturing the downstream performance. 

\textbf{Usage}: Essentially, an average of the subgroup AUCs that helps in easier comparison of different debiasing settings by having to follow only one metric. A lower GMB AUC like subgroup-specific AUC is a good indicator of unintended bias in the toxicity detection pipeline against one or more groups. 

\item \textbf{Pinned Bias ($\mathbf{pB}$):} In a work parallel to \cite{borkanmetrics2019}, \citet{paper21} introduced a family of bias evaluation metrics, pinned Bias ($pB$):
\begin{equation}
pB_T = \sum_{t\in T}{\frac{|p(``toxic"|t) - \phi|}{|T|}}
\end{equation}
where (i) $p(``toxic"|t)$ is the prediction probability of the toxicity label for a sentence containing only  term $t$, (ii) $\phi$ is the pinned/threshold value that differs for different members of the metric family. For example, to penalize $p(``toxic"|t)$ values above 0.5, $\phi = min(p(``toxic"|t),0.5)$ can be used. Here the pinning is on the threshold probability that depicts how much toxicity can be allowed for a term. The limitation of above formulation is that $\phi$ is constant for all terms in $T$, while in real world the thresholds are dynamic.

\textbf{Usage:} pB can be used in tandem or even as a replacement for pinned AUC. While pAUCED requires the creation of a balanced auxiliary dataset for each term $t \in BSW$, pB utilizes the same dataset and is easier to measure.

\end{itemize}

\section{Case Study on Knowledge-drift}
\textbf{Lexical debiasing.} To overcome lexical bias (Section \ref{sec:lexicalbias}), \citet{paper21} employed a knowledge-based generalisation method. It involved the replacement of the BSWs in the training dataset with an ancestor from the WordNet \cite{wordnet} hypernym-tree. 

\textbf{Knowledge-drift hypothesis.} Motivated by the energy-bias analogy (refer to Section \ref{sec:takeways}), we conjecture that replacing all occurrences $w\in BSW$ with its wordnet ancestor $a\in A$ will shift the bias from $w$ to $a$. While generalization can be true for all the debiasing methods we discussed in this survey, for this experiment, we focus on the wordnet-based substitution method for lexical 
\label{sec:casestudybiasshift}debiasing. 

\textbf{Limited evaluation.} Post substitution of $W$ with $A$, the authors employed the BSW on the original $W$ to evaluate bias. There is no discussion on evaluating the bias on terms in $A$, which is now a substitute for $W$.

\textbf{Experimental setup.} All experiments were carried out on a Ubuntu 18.04.5 LTS system with $126G$ RAM and $32G$ Tesla V100. We apply debiasing method on the W\&H dataset \cite{waseemandhovy} to verify the hypothesis and compare by training a BERTweet \cite{bertweet} classifier. 

\begin{table}[!h]
\caption{(a): BSWs $w$ with their corresponding selected generalisations $a$ according to the Wordnet-3 scheme  \cite{paper21}. (b): $pB$ values for the set of BSWs $W$ and $A$ for (i) the model obtained on original W\&H dataset $M_{bias}$ and (ii) the model obtained after lexical database generalisation $M_{gen}$.}
\label{tab:casestudybiasshift}
\subfloat[]{
 \begin{tabular}{ | p{10cm} | c | }
    \hline
    \multicolumn{1}{|c|}{$w\in W$} &$a \in A$\\
    \hline
    muslim, prophet, woman, christian, girl, terrorist, slave, man, child, driver&being\\
    \hline
    feminist, civilian, liar, comedian, god &someone\\
    \hline
    hate, slavery, hatred, want, truth, freedom&state\\
    \hline
\end{tabular}
}
\subfloat[]{
\begin{tabular}{|c | c | c |}
    \hline
    Metric&$M_{bias}$&$M_{gen}$\\
    \hline
    $pB_{W}$&$0.027$&$0$\\
    $pB_{A}$&$0.002$&$0.016$\\
    \hline
 \end{tabular}
}
\vspace{-3mm}
\end{table}
\textbf{Observation.} Table \ref{tab:casestudybiasshift}(a) lists the replaced words and their respective generalizations. The observations in Table \ref{tab:casestudybiasshift}(b) indicate a shift of bias from source to target words. Since the terms in the set $A$ are more likely to be present in non-toxic comments, this shift in bias can also be detrimental to the non-toxical class. While this is a proof of concept case study, it is expected that the other mitigation techniques by \citet{paper21} based on knowledge-based generalizations suffer from a similar shift of bias instead of actually debiasing the dataset.

\textbf{Resolving bias shift.} Note that this shift of lexical bias is towards a more general set of terms $A$, where debiasing on this generalized dataset through upsampling \cite{paper20} directly from an OSN (such as Twitter) seems like a feasible next step. For example, newly scraped comments containing terms such as \textit{muslim} can not be directly added to the training dataset. It is because the number of toxic comments can be expected to be higher. However, if \textit{muslim} is first generalized to \textit{being}, randomly scraping new comments with this new keyword can lead to a significantly lower toxicity ratio.

\section{Reproducibility of debiasing Methods}
\label{sec:rep_research}
\textbf{Motivation.} In this section we reproduce some of the bias mitigation methods described in the respective sections of bias. We mainly extend upon the results reported by \citet{paper10} and \citet{paper9} to incorporate W\&H \cite{waseemandhovy} and Davidson \cite{davidsondataset} datasets along with Founta \cite{founta}. By reproducing these methods, we can comment not just on the theoretical discussion of the debiasing techniques but also on their engineering feasibility. We hope to give the readers a glimpse of various settings (dataset, embedding, evaluation combination) for the debiasing methods  presented in this survey.

\textbf{Experimental Setup.} All experiments were carried out on a Ubuntu 18.04.5 LTS system with $126G$ RAM and $32G$ Tesla V100. The hyper-parameters and epochs were kept intact as provided in the respective methodology. 

\begin{table}[!h]
\caption{Reproduced results for racial debiasing as experimented by \citet{paper10}. (a) Results on W\&H \cite{waseemandhovy}, (b) Results on Davidson \cite{davidsondataset}, (c) Results on Founta \cite{founta}.}
\label{tab:racial_bias}
\subfloat[]{
\scalebox{0.67}{
\begin{tabular}{|l|l|l|l|l|}
\hline
Method/Result & W FPR & B FPR & W \% & B \% \\\hline
Random        & $0.072$ & $0.133$ & $0.267$  & $0.481$  \\\hline
Original      & $0.025$ & $0.024$ & $ 0.162$  & $0.380$  \\\hline
LearnedMixIn    & $ 0.028$ & $0.024$ & $0.178$ & $0.405$ \\\hline
\end{tabular}
}
\newline
}
\subfloat[]{
\scalebox{0.67}{
\begin{tabular}{|l|l|l|l|l|}
\hline
Method/Result & W FPR & B FPR & W \% & B \% \\\hline
Random        & $0.029$ & $0.022$ & $0.045$  & $0.033$  \\\hline
Original      & $0.032$ & $0.009$ & $ 0.063$  & $0.015$  \\\hline
LearnedMixIn    & $0.193$ & $0.308$ & $0.194$ & $0.313$ \\\hline
\end{tabular}
}
}
\subfloat[]{
\scalebox{0.67}{
\begin{tabular}{|l|l|l|l|l|}
\hline
Method/Result & W FPR & B FPR & W \% & B \% \\\hline
Random        & $0.017$ & $0.058$ & $0.303$  & $0.719$  \\\hline
Original      & $0.040$ & $0.115$ & $0.300$  & $0.725$  \\\hline
Datamap-Easy  & $0.018$ & $0.053$ & $0.302$  & $0.717$  \\\hline
Datamap-Ambi  & $0.020$ & $0.077$ & $0.307$  & $0.725$  \\\hline
Datamap-Hard  & $0.020$ & $0.073$ & $0.287$  & $0.716$  \\\hline
Aflite        & $0.017$ & $0.058$ & $0.303$  & $0.719$ \\\hline
LearnedMixIn    & $0.0275$ & $0.075$ & $0.281$ & $0.694$ \\\hline
\end{tabular}
}
}
\vspace{-7mm}
\end{table}

\subsection{Racial Bias}
For studying racial bias, we reproduce the results reported in \citet{paper10}. As observed in Table \ref{tab:racial_bias}(c) for the large scale dataset (like Founta \cite{founta}), data filtering methods like Aflite that eventually operate on only $33\%$ of the dataset provide a comparable drop (less is more) in $FPR$. Meanwhile, the LearnedMixIn method \cite{clarklearnedmixin} that adopts model regularisation provides the smallest percentage of Black tweets samples wrongly labelled as toxic. For W\&H and Davidson datasets applying data filtering and utilising only $33\%$ of the dataset would leave us only $\approx5k$ and $\approx8k$ tweets, respectively. Hence we did not test data filtering on small scale datasets and only reported unexpected and LearnMixIn results. Debiasing in both datasets via LearnMixIn (Tables \ref{tab:racial_bias}(a) and \ref{tab:racial_bias}(b)) is not better than the original dataset composition. This we assume could be happening due to the small volume of samples and the presence of other biases (lexical, topical) in the datasets.

\subsection{Gender Bias}
Here we reproduce the results from \citet{paper9} by testing the combination of embedding and model against an exhaustive set of gender debiasing techniques. For the transfer learning, we test on Waseem and Davidson using Founta for training. In line with existing literature, we observe in Table \ref{tab:gender_1} that pre-trained word embedding, on average, gives higher FPED and FNED due to their intrinsic gender bias. Similarly, as observed from Table \ref{tab:gender_2} transfer learning leads to a higher F1 score, but it comes at the cost of increasing FPED and FNED. A combination of debiased word2vec, gender-swapping and transfer learning while a more complex set seems to be the best tradeoff between reducing F1 and increasing FPED-FNED values. However, no one setting is best across datasets. 
\begin{table}[!h]
\caption{Reproduced results for gender debiasing employing CNN, GRU and $\alpha$-GRU on $3$ different embeddings -- $R$: randomly initialised, $F$: fasttext, $W$: word2vec.  (a) Results on W\&H \cite{waseemandhovy}, (b) Results on Davidson \cite{davidsondataset}, (c) Results on Founta \cite{founta}.}
\label{tab:gender_1}
\subfloat[]{
\scalebox{0.67}{
\begin{tabular}{|p{1cm}|p{0.65cm}|l|l|l|l|l|}
\hline
Method                        & Emd. & FPR   & FNR   & FPED & FNED & F1 \\\hline
\multirow{3}{*}{CNN}          & R    & $0.039$ & $0.331$ & $0.067$& $0.586$ & $0.885$    \\\cline{2-7}
                              & F  & $0.060$ & $0.213$ & $0.097$ & $0.384$ &$0.902$    \\\cline{2-7} 
                              & W  & $0.053$ & $0.247$ &$0.088$      &$0.445$      &$0.899$    \\\hline 
\multirow{3}{*}{GRU}          & R    & $0.076$ & $0.258$ &$0.134$ &$0.439$      &$0.879$    \\\cline{2-7}
                              & F  & $0.071$ & $0.277$ &$0.120$      &$0.477$      & $0.877$    \\\cline{2-7} 
                              & W  & $0.082$ & $0.227$ &$0.141$&$0.400$      & $0.882$    \\\hline
\multirow{3}{*}{$\alpha$-GRU} & R    & $0.043$ & $0.243$ &$0.073$      &$0.426$      &$0.907$    \\\cline{2-7}
                              & F  & $0.080$ & $0.228$ &$0.142$       &$0.414$      & $0.885$   \\\cline{2-7}
                              & W  & $0.080$ & $0.202$ & $0.142$     &$0.365$       &$0.892$   \\\hline
\end{tabular}
}
}
\subfloat[]{
\scalebox{0.67}{
\begin{tabular}{|p{1cm}|p{0.65cm}|l|l|l|l|l|}
\hline
Method                        & Emd. & FPR   & FNR   & FPED & FNED & F1 \\\hline
\multirow{3}{*}{CNN}          & R    & 0.010 & 0.842 & 0.019 & 1.561 & 0.917 \\ \cline{2-7} 
                              & F  & 0.013 & 0.762 & 0.025 & 1.415 & 0.926 \\ \cline{2-7} 
                              & W  & 0.008 & 0.768 & 0.016 & 1.439 & 0.928 \\ \hline
\multirow{3}{*}{GRU}          & R    & 0.043 & 0.654 & 0.082 & 1.222 & 0.916 \\ \cline{2-7} 
                              & F  & 0.059 & 0.624 & 0.108 & 1.160 & 0.908 \\ \cline{2-7} 
                              & W  & 0.035 & 0.735 & 0.065 & 1.370 & 0.914 \\ \hline
\multirow{3}{*}{$\alpha$-GRU} & R    & 0.037 & 0.750 & 0.069 & 1.372 & 0.912 \\ \cline{2-7} 
                              & F  & 0.041 & 0.695 & 0.074 & 1.274 & 0.914 \\ \cline{2-7} 
                              & W  & 0.047 & 0.683 & 0.087 & 1.256 & 0.910 \\ \hline
\end{tabular}
}
}
\subfloat[]{
\scalebox{0.67}{
\begin{tabular}{|p{1cm}|p{0.65cm}|l|l|l|l|l|}
\hline
Method                        & Emd. & FPR   & FNR   & FPED & FNED & F1 \\\hline
\multirow{3}{*}{CNN}          & R    & 0.04  & 0.327 & 0.071 & 0.586 & 0.883 \\ \cline{2-7} 
                              & F  & 0.048 & 0.247 & 0.077 & 0.449 & 0.902 \\ \cline{2-7} 
                              & W  & 0.044 & 0.262 & 0.069 & 0.487 & 0.901 \\ \hline
\multirow{3}{*}{GRU}          & R    & 0.143   &  0.325   &  0.254   & 0.560 & 0.786 \\ \cline{2-7} 
                              & F  & 0.070   & 0.125  & 0.125   & 0.216 & 0.909 \\ \cline{2-7} 
                              & W  & 0.097  & 0.119  & 0.174   & 0.203 & 0.89 \\ \hline
\multirow{3}{*}{$\alpha$-GRU} & R    & 0.081 & 0.232 & 0.142 & 0.422 & 0.883 \\ \cline{2-7} 
                              & F  & 0.098 & 0.183 & 0.170 & 0.327 & 0.884 \\ \cline{2-7} 
                              & W  & 0.081 & 0.202 & 0.138 & 0.357 & 0.891 \\ \hline
\end{tabular}
}
}
\end{table}
\begin{table}
\caption{Reproduced results for gender debiasing employing CNN, GRU and $\alpha$-GRU on a combination of $3$ debiasing techniques -- $DE$: debiased word2vec, $GS$: gender swapping and  $FT$: transfer learning by training on Founta \cite{founta}. \cmark marks the combination of debiasing used. (a) Results on W\&H \cite{waseemandhovy}, (b) Results on Davidson \cite{davidsondataset}.}
\label{tab:gender_2}
\subfloat[]{
\scalebox{0.67}{
\begin{tabular}{|l|l|l|l|l|l|l|l|l|}
\hline
Method                  & DE & GS & FT & FPR    & FNR   & FPED  & FNED  & F1    \\ \hline
\multirow{7}{*}{CNN}          & \cmark  & -  & -  & 0.0294 & 0.316 & 0.045 & 0.567 & 0.897 \\ \cline{2-9} 
                              & -  &\cmark & -  & 0.047  & 0.297 & 0.080 & 0.555 & 0.890 \\ \cline{2-9} 
                              & -  & -  & \cmark  & 0.054  & 0.331 & 0.090 & 0.600 & 0.875 \\ \cline{2-9} 
                              & \cmark  & \cmark  & -  & 0.054  & 0.267 & 0.087 & 0.491 & 0.893 \\ \cline{2-9} 
                              & -  & \cmark  & \cmark  & 0.118 & 0.909 & 0.218 & 1.422 & 0.644 \\ \cline{2-9} 
                              & \cmark & -  & \cmark  & 0.124  & 0.871 & 0.223 & 1.365 & 0.656 \\ \cline{2-9} 
                              & \cmark  & \cmark  & \cmark  & 0.118  & 0.909 & 0.218 & 1.422 & 0.644 \\ \hline
\multirow{7}{*}{GRU}          & \cmark  & -  & -  & 0.070  & 0.297 & 0.123 & 0.515 & 0.872 \\ \cline{2-9} 
                              & -  & \cmark  & -  & 0.076  & 0.247 & 0.130 & 0.427 & 0.882 \\ \cline{2-9} 
                              & -  & -  & \cmark & 0.095  & 0.427 & 0.166 & 0.723 & 0.818 \\ \cline{2-9} 
                              & \cmark  & \cmark  & -  & 0.048  & 0.342 & 0.084 & 0.581 & 0.874 \\ \cline{2-9} 
                              & -  & \cmark  & \cmark  & 0.134  & 0.816 & 0.252 & 1.27  & 0.668 \\ \cline{2-9} 
                              & \cmark  & -  & \cmark  & 0.064  & 0.712 & 0.118 & 1.185 & 0.746 \\ \cline{2-9} 
                              & \cmark  & \cmark  & \cmark  & 0.065  & 0.708 & 0.119 & 1.177 & 0.746 \\ \hline
\multirow{7}{*}{$\alpha$-GRU} & \cmark  & -  & -  & 0.065  & 0.297 & 0.115 & 0.525 & 0.877 \\ \cline{2-9} 
                              & -  & \cmark  & -  & 0.075  & 0.255 & 0.130 & 0.456 & 0.881 \\ \cline{2-9} 
                              & -  & -  & \cmark  & 0.061  & 0.278 & 0.107 & 0.487 & 0.885 \\ \cline{2-9} 
                              & \cmark  & \cmark  & -  & 0.097  & 0.232 & 0.169 & 0.400 & 0.872 \\ \cline{2-9} 
                              & -  & \cmark  & \cmark  & 0.071  & 0.247 & 0.132 & 0.426 & 0.886 \\ \cline{2-9} 
                              & \cmark  & -  & \cmark  & 0.085 & 0.243 & 0.148 & 0.445 & 0.877 \\ \cline{2-9} 
                              & \cmark  & \cmark  & \cmark  & 0.092  & 0.247 & 0.155 & 0.437 & 0.871 \\ \hline
\end{tabular}
}
}
\subfloat[]{
\scalebox{0.67}{
\begin{tabular}{|l|l|l|l|l|l|l|l|l|}
\hline
Method                  & DE & GS & FT & FPR    & FNR   & FPED   & FNED  & F1    \\ \hline
\multirow{7}{*}{CNN}          & \cmark  & -  & -  & 0.446  & 0.537 & 0.791  & 1.0   & 0.658 \\ \cline{2-9} 
                              & -  & \cmark  & -  & 0.008  & 0.793 & 0.016  & 1.476 & 0.925 \\ \cline{2-9} 
                              & -  & -  & \cmark  & 0.037  & 0.098 & 0.056  & 0.149 & 0.940 \\ \cline{2-9} 
                              & \cmark  & \cmark  & -  & 0.008  & 0.860 & 0.015  & 1.567 & 0.917 \\ \cline{2-9} 
                              & -  & \cmark  & \cmark  & 0.085  & 0.927 & 0.148  & 1.707 & 0.867 \\ \cline{2-9} 
                              & \cmark  & -  & \cmark  & 0.235  & 0.774 & 0.424  & 1.446 & 0.792 \\ \cline{2-9} 
                              & \cmark  & \cmark & \cmark  & 0.129  & 0.860 & 0.224  & 1.580 & 0.849 \\ \hline
\multirow{7}{*}{GRU}          & \cmark  & -  & -  & 0.0415 & 0.685 & 0.075  & 1.278 & 0.914 \\ \cline{2-9} 
                              & -  & \cmark  & -  & 0.035  & 0.735 & 0.065  & 1.370 & 0.914 \\ \cline{2-9} 
                              & -  & -  & \cmark  & 0.491  & 0.501 & 0.712  & 1.0   & 0.679 \\ \cline{2-9} 
                              & \cmark  & \cmark  & -  & 0.0468 & 0.716 & 0.083  & 1.29  & 0.908 \\ \cline{2-9} 
                              & -  & \cmark  & \cmark  & 0.009  & 1.0   & 0.014  & 1.821 & 0.898 \\ \cline{2-9} 
                              & \cmark  & -  & \cmark  & 0.003  & 1.0   & 0.004  & 1.846 & 0.901 \\ \cline{2-9} 
                              & \cmark  & \cmark  & \cmark  & 0.006  & 0.932 & 0.011  & 1.698 & 0.909 \\ \hline
\multirow{7}{*}{$\alpha$-GRU} & \cmark  & -  & -  & 0.034  & 0.762 & 0.0618 & 1.409 & 0.912 \\ \cline{2-9} 
                              & -  & \cmark  & -  & 0.031  & 0.738 & 0.057  & 1.366 & 0.916 \\ \cline{2-9} 
                              & -  & -  & \cmark  & 0.079  & 0.120 & 0.123  & 0.182 & 0.906 \\ \cline{2-9} 
                              & \cmark  & \cmark  & -  & 0.035  & 0.781 & 0.065  & 1.402 & 0.909 \\ \cline{2-9} 
                              & -  & \cmark & \cmark  & 0.041  & 0.762 & 0.075  & 1.378 & 0.907 \\ \cline{2-9} 
                              & \cmark  & -  & \cmark  & 0.038  & 0.787 & 0.068  & 1.439 & 0.907 \\ \cline{2-9} 
                              & \cmark  & \cmark  & \cmark  & 0.014  & 0.860 & 0.024  & 1.555 & 0.914 \\ \hline
\end{tabular}
}
}
\end{table}
\end{document}